\documentclass[letterpaper,12pt]{article}
\usepackage{color,amssymb,enumerate,float}

\usepackage{ifpdf}
\ifpdf
	\usepackage[pdftex]{graphicx}
	\usepackage[pdftex,unicode,implicit]{hyperref}

	\hypersetup{
  	pdftitle     = {quantizingkc}, 
  	pdfkeywords  = {},
  	pdfauthor    = {},
  	pdfcreator   = {pdf\LaTeXe\ with package \flqq hyperref\frqq},
  	pdfproducer  = {pdf\LaTeXe\ with package \flqq hyperref\frqq},
  	pdfpagemode  = UseNone,  
  	pdffitwindow = true,  
  	unicode      = true,
  	plainpages   = true,
  	colorlinks   = true,  
  	citecolor    = black,  
  	urlcolor     = blue, 
  	linkcolor    = black
	}

\else

  \usepackage[dvips]{graphicx}

\fi    

\makeatletter
\@addtoreset{equation}{section}
\makeatother

\begin{document}
	
\thispagestyle{empty}

\begin{center}

{\bf \LARGE Quantization of the Ho\v{r}ava theory at the kinetic-conformal point}
\vspace*{15mm}

{\large Jorge Bellor\'{\i}n}$^1$
{\large and Alvaro Restuccia}$^2$
\vspace{3ex}

{\it Department of Physics, Universidad de Antofagasta, 1240000 Antofagasta, Chile.}
{\it Department of Physics, Universidad Sim\'on Bol\'{\i}var, 1080-A Caracas, Venezuela.} 
\vspace{3ex}

$^1${\tt jbellori@gmail.com,} \hspace{1em}
$^2${\tt arestu@usb.ve}

\vspace*{15mm}
{\bf Abstract}
\begin{quotation}{\small
The Ho\v{r}ava theory depends on several coupling constants. The kinetic term of its Lagrangian depends on one dimensionless coupling constant $\lambda$. For the particular value $\lambda = 1/3$ the kinetic term becomes conformal invariant, although \emph{the full Lagrangian does not have this symmetry}. For any value of $\lambda$ the nonprojectable version of the theory has second-class constraints which play a central role in the process of quantization. Here we study the complete nonprojectable theory, including the Blas-Pujol\`as-Sibiryakov interacting terms, at the kinetic-conformal point $\lambda = 1/3$. The generic counting of degrees of freedom indicates that this theory propagates the same physical degrees of freedom of General Relativity. We analyze this point rigorously taking into account all the $z=1,2,3$ terms that contribute to the action describing quadratic perturbations around the Minkowski spacetime. We show that the constraints of the theory and equations determining the Lagrange multipliers are strongly elliptic partial differential equations, an essential condition for a constrained phase-space structure in field theory. We show how their solutions lead to the two independent tensorial physical modes propagated by the theory. We also obtain the reduced Hamiltonian. These arguments strengthen the consistency of the theory. We find the restrictions on the space of coupling constants to ensure the positiveness of the reduced Hamiltonian. We obtain the propagator of the physical modes, showing that there are not ghosts and that the propagator effectively acquires the $z=3$ scaling for all physical degrees of freedom at the high energy regime. By evaluating the superficial degree of divergence, taking into account the second-class constraints, we show that the theory is power-counting renormalizable. we analyse, in the path integral formulation of the theory, the measure associated to the second-class constraints both in the canonical and the Lagrangian (FDiff-covariant) formalisms.
}
\end{quotation}

\end{center}

\thispagestyle{empty}

\newpage
\section{Introduction}
Ho\v{r}ava theory is a proposition of a perturbatively renormalizable and unitary quantum field theory of gravity in $3+1$ spacetime dimensions (although the general principles can be applied to other dimensions as well). The original formulation was done in Ref.~\cite{Horava:2009uw}, with related concepts previously considered in Ref.~\cite{Horava:2008ih}. The main idea is to break the relativistic symmetry (at least in the gravitational sector) by introducing a timelike direction into the spacetime, with absolute physical meaning, with the hope of obtaining a renormalizable theory. The spacetime is foliated in terms of spacelike hypersurfaces along this direction. The allowed coordinate transformations, instead of the general transformations between time and space characteristic of general relativity (GR), are the ones that preserve the given foliation. The gauge symmetry group of the theory is then the foliation-preserving diffeomorphisms group (FDiff). A FDiff-covariant Lagrangian allows the inclusion of interacting terms with higher order spatial derivatives of the metric field (which is dimensionless), without the need of increasing the order in time derivatives. Thus, the central aim is that the higher spatial curvature terms that contribute to the propagators improve the renormalization properties of the theory while keeping under control the number of poles since no higher time derivatives are added. This program is a reminiscent of the relativistic higher curvature theories. However, the crucial difference is that in the latter theories, in order to preserve the relativistic symmetry, the order of the time derivatives must be increased as higher curvature terms are included. Among the added poles there arise ghosts that break the unitarity of the theory \cite{Stelle:1976gc}.

Since its original formulation in \cite{Horava:2009uw}, the theory has evolved in several directions. Initially the potential was restricted by the so-called detailed balance principle, which basically postulates that the potential of the $3+1$ theory must be derived from a purely spatial three-dimensional Lagrangian. Currently many authors prefer to abandon this principle and instead consider the general, potentially renormalizable theory that includes in the potential all the terms compatible with the FDiff gauge symmetry. Besides this, the theory has two separate main versions, the projectable and the nonprojectable versions. These two ways of formulating the theory, already studied in \cite{Horava:2009uw}, are characterized by the lapse function being a function only of the time coordinate (projectable version) or a general function of time and space (nonprojectable version). Among other developments, the projectable version has been modified by including an extra $U(1)$ gauge symmetry \cite{Horava:2010zj}, eliminating in this way the extra degree of freedom. On the nonprojectable side, a wide class of interacting terms compatible with the FDiff symmetry was incorporated in Ref.~\cite{Blas:2009qj}. These terms make the potential dependent on the lapse function $N$ via the FDiff-covariant vector $a_i = \partial_i \ln N$.  Following the spirit of renormalizable gauge theories, the Lagrangian should include all the terms, up to the order required for renormalization, compatible with the underlying gauge symmetry. We refer as the nonprojectable Ho\v{r}ava theory to the theory including the $a_i$ terms. An $U(1)$ extension similar to the one of the projectable case was proposed for the nonprojectable version in \cite{Zhu:2011xe}. The truncation of the nonprojectable theory to second order in derivatives has been found \cite{Blas:2009ck,Jacobson:2010mx,Jacobson:2013xta} to be related to the Einstein-aether theory \cite{Jacobson:2000xp}; specifically the solutions of the latter having a hypersurface-orthogonal aether vector are solutions of the former (but the converse is not true in general \cite{Jacobson:2013xta}). Recently the Ho\v{r}ava theory, both in the projectable and the nonprojectable versions, has been reproduced by gauging (making dynamical) the Newton-Cartan geometry \cite{Hartong:2015zia}.

In the nonprojectable case, including the $a_i$ terms of \cite{Blas:2009qj}, the closure of the algebra of constraints of the classical Hamiltonian formulation has been shown \cite{varios:hamiltoniannokc} (see also \cite{Bellorin:2012di}). There the crucial role of the $a_i$ terms in improving the structure of the constraints was noticed. Indeed, one of the motivations of \cite{Blas:2009qj} to include these terms was to improve the mathematical structure of the field equations in the Lagrangian scheme. Implicitly assuming the invertibility of the Legendre transformation, in the Hamiltonian analysis of Refs.~\cite{varios:hamiltoniannokc}, the presence of an extra degree of freedom was corroborated. The extra mode was previously identified in Ref.~\cite{Blas:2009qj} with a well-behaved dispertion relation (under suitable restrictions on the space of parameters). Among several features that have been studied for the extra mode, it has been found that, whenever one forces the kinetic term to adopt the relativistic version at low energies, it suffers from the so-called strong-coupling problem \cite{varios:strongcoupling}. A feasible resolution of this problem is to demand that the scale of activation of higher order operators is low enough \cite{Blas:2009ck}.

In Ref.~\cite{Bellorin:2013zbp} the case in which the invertibility of the Legendre transformation does not hold was analyzed. This happens when the independent (dimensionless) coupling constant arising in the kinetic term, $\lambda$, acquires the specific value $\lambda = 1/3$. At this value the kinetic term acquires a conformal invariance \cite{Horava:2009uw}, but \emph{the whole theory is not conformally invariant} since in general the terms in the potential break the conformal symmetry (unless only specific terms, like $(\mbox{Cotton})^2$, are included in the potential such that it is rendered conformally invariant). For this reason we call the point $\lambda = 1/3$ the kinetic-conformal (KC) point, and use the same name for the Ho\v{r}ava theory formulated at this point.

At the KC point there arise two extra second-class constraints \cite{Bellorin:2013zbp}. Qualitatively, one may regard the presence of these new constraints as a consequence of the lack of invertibility of the Legendre transformation at the KC point. The two constraints eliminate precisely the extra mode. We consider this a very interesting property, since the number of degrees of freedom of the KC Ho\v{r}ava theory coincides with the one of GR (the $U(1)$ extensions also eliminate the extra mode \cite{Horava:2010zj,Zhu:2011xe}). In Ref.~\cite{Bellorin:2013zbp} the closure of the algebra of constraints assuming a general, unspecified, potential was shown. In addition, a model with soft breaking of the conformal invariance was considered there, corroborating the consistent structure of constraints and conditions for the Lagrange multipliers with explicit equations. Moreover the perturbative version of the effective large-distance action of the KC theory at quadratic order in perturbations is physically equivalent to perturbative GR.

We devote this paper to deepening the features of the nonprojectable Ho\v{r}ava theory at the KC point. We pose ourselves two main objectives. The first one is to further advance the knowledge of the classical Hamiltonian formulation, which is fundamental for the consistency of the theory. We would like to get explicit expressions for all the constraints and conditions for the Lagrange multipliers when the potential contains all the possible interacting terms up to $z=3$, the minimal order to get renormalizability in $3+1$ spacetime dimensions. To this end we adopt a perturbative approach, taking in the potential all the terms that contribute to the quadratic action.

Our second objective is to enter into the process of quantization of the KC Ho\v{r}ava theory. From the results in the linearized classical theory we obtain the reduced Hamiltonian and study the conditions needed to guarantee the positiveness of its spectrum. Then we study the propagator of the physical modes. Getting explicitly the independent propagators is one of the first tasks to do in the Ho\v{r}ava theory since in this way one elucidates if the theory really possesses the ultraviolet (UV) improved and ghost-free propagators heuristically proposed in the original paper of Ho\v{r}ava \cite{Horava:2009uw}. Indeed, without the KC condition, there is a sector of the space of parameters where the extra mode becomes a ghost \cite{Blas:2009qj}. Another counterexample is that in the theory with detailed balance the operator with the highest derivative does not contribute to the propagator of the extra mode.

On the basis of the physical propagators, we give arguments on the power-counting renormalizability of the theory, specifically by computing the superficial degree of divergence of one-particle-irreducible (1PI) diagrams. Our interest is in evaluating the power of divergences directly on the gravitational variables. This is more acute than, for example, using toy models like scalar-field theories since in these models precisely the constraints are not represented.

Another question we address about the quantization of the theory is what happens when it is formulated in the nonreduced phase space, as it is typically done in gauge theories. Here the main point is that the nonprojectable Ho\v{r}ava theory, with or without the KC condition, \emph{has second-class constraints}. Whenever these constraints are not solved, which is by definition the formulation in the nonreduced phase space, one is forced to take into account their second-class nature under any scheme of quantization. In this work we study the path-integral quantization, where the presence of second-class constraints requires the modification of the measure. We consider both the Hamiltonian and the Lagrangian (FDiff-covariant) formulation of the path integral. In particular it is important to conciliate the Lagrangian path integral with the canonical one since if one starts solely with the Lagrangian formulation then one does not know the correct measure associated to the second-class constraints.

Several authors have made computations in the quantized Ho\v{r}ava gravity or in related toy models without the KC condition. Among them, power-counting renormalizability criteria have been proposed in \cite{Visser:2009fg,Visser:2009ys} (actually, these papers provide a general framework applicable to the KC case). The propagator for a nonprojectable model with $z=1$ and $z=3$ terms was studied in Ref.~\cite{Pospelov:2010mp}. In that paper several considerations about the bounds imposed by the coupling to matter, where the experimental restrictions on Lorentz violations are very strong, were considered. In Refs.~\cite{varios:stochastic} the renormalization of the projectable theory with detailed balance was considered with the methods of stochastic quantization. The one-loop renormalization of the conformal reduction of the projectable theory in $2+1$ dimensions was analyzed in \cite{Benedetti:2013pya}. Gaussian and non-Gaussian fixed points in the renormalization flow as well as their consequences on asymptotic freedom and asymptotic safety have been investigated in the projectable Ho\v{r}ava theory and its couplings in Refs.~\cite{varios:asymptoticfreedom}. The power-counting renormalizability of models with mixed time and spatial derivative terms has been considered in Refs.~\cite{Colombo:2014lta,Colombo:2015yha}. Recently, the authors of Ref.~\cite{Barvinsky:2015kil} showed the complete perturbative renormalizability of the projectable theory (without detailed balance). To this end they used nonlocal gauge-fixing conditions. The quantization of Ho\v{r}ava theory has also been connected to causal dynamical triangulations \cite{varios:cdt}.

This paper is organized as follows: in Section 2 we study the consistency of the classical Hamiltonian formulation. We first present the general results for the Hamiltonian formulation with an unspecified potential. Then we address the solutions of the constraints in a perturbative approach. In Section 3 we perform the quantum computations. This section is divided in three parts. In the first one we study the reduced Hamiltonian and the positiveness of its spectrum. In the second one we present the propagator of the physical modes and consider power-counting renormalizability. In the last one we study the path integral in the nonreduced phase space. We devote Section 4 to highlighting the fact that the nonprojectable theory without the KC condition also has second-class constraints and the measure is affected by them. Finally, we present some discussion and conclusions about our results. There is also some appended material relevant for the themes discussed in this paper.

\section{Consistency of the classical Hamiltonian}
\subsection{The general canonical theory}
The formulation of the theory starts with the assumption that in the spacetime there is a timelike direction and a foliation in terms of spacelike hypersurfaces along it with absolute physical meaning. The underlying symmetry of the theory is not the set of general coordinate transformations between time and space but the restricted set of coordinate transformations that do not change the absolute timelike direction and its associated foliation. Thus, the gauge symmetry group is the group of diffeomorphisms over the spacetime that preserve the given foliation (FDiff) \cite{Horava:2009uw}. Its action on the coordinates $(t,\vec{x})$ is
\begin{equation}
 \delta t = f(t) \,,
\hspace{2em}
 \delta x^i = \zeta(t,\vec{x}) \,.
\end{equation}
The gravitational part of the theory is formulated in the Arnowitt-Deser-Misner (ADM) variables, $g_{ij}$, $N$ and $N_i$. Under FDiff these variables transform as
\begin{equation}
\begin{array}{l}
\delta N = \zeta^k \partial_k N + f \dot{N} + \dot{f} N \,,
\\[1ex]
\delta N_i = \zeta^k \partial_k N_i + N_k \partial_i \zeta^k 
+ \dot{\zeta}^j g_{ij} + f \dot{N}_i + \dot{f} N_i \,,
\\[1ex]
\delta g_{ij} = \zeta^k \partial_k g_{ij} + 2 g_{k(i} \partial_{j)} \zeta^k 
+ f \dot{g}_{ij}  \,,
\end{array}                
\label{fdiff}
\end{equation}
where the dot denotes time derivative, $\dot{N} = \frac{\partial N}{\partial t}$. The action of the FDiff group allows two different formulations of the theory, each one characterized by the kind of dependence the lapse function $N$ has. In one version, called the projectable version, $N$ is a function of only the time and this condition is preserved by FDiff (which can be deduced from (\ref{fdiff})). The other version, in which $N$ depends both in time and space, is called the nonprojectable case. The theory we study in this paper belongs to the nonprojectable case. In this case the Hamiltonian constraint is present as a local constraint, like in GR. On the other hand, due to the reduced symmetry group, the behavior of the Hamiltonian constraint is different to GR.

With the aim of getting renormalizability while avoiding unitarity loss, the theory is designed in such a way that at high energies it should naturally exhibit an anisotropic scaling between time and space,
\begin{equation}
t \rightarrow b^z t \,,
\hspace{2em}
\vec{x} \rightarrow b \vec{x} \,.
\end{equation}
The parameter $z$ characterizes the degree of anisotropy. Power-counting arguments lead us to consider $z=3$ in $3+1$ spacetime dimensions as the minimal degree of anisotropy to get a renormalizable theory \cite{Horava:2009uw}. Under this scenario the dimensionality (in momentum powers) of the coordinates and field variables is postulated as \cite{Horava:2009uw}
\begin{equation}
  [\,t\,] = - z \,, \hspace{1.5em} 
  [\,\vec{x}\,] = - 1 \,, \hspace{1.5em}
  [\,g_{ij}\,] = [\,N\,] = 0 \,, \hspace{1.5em}
  [\,N_i\,] = z - 1 
\label{dimensions}
\end{equation}
(for the intrinsic formulation of the quantum theory it is not essential to have the structure of a four-dimensional spacetime metric, but in any case it can be recovered by a suitable rescaling of the time coordinate using an emerging light-speed constant \cite{Horava:2009uw}).

The action of the complete nonprojectable theory is \cite{Horava:2009uw,Blas:2009qj}
\begin{equation}
 S = \int dt d^3x \sqrt{g} N 
       \left( \frac{1}{2\kappa} G^{ijkl} K_{ij} K_{kl} - \mathcal{V} \right),
\label{lagrangianaction}
\end{equation}
where
\begin{eqnarray}
K_{ij} & = & \frac{1}{2N} ( \dot{g}_{ij} - 2 \nabla_{(i} N_{j)} ) \,,
\\[1ex]
G^{ijkl} & = &
\frac{1}{2} \left( g^{ik} g^{jl} + g^{il} g^{jk} \right) 
- \lambda g^{ij} g^{kl} 
\end{eqnarray}
and $\lambda$ is a dimensionless constant. Two comments are in order: first, if $z = d = 3$, $\kappa$ becomes a dimensionless coupling constant \cite{Horava:2009uw}. Second, in a relativistic theory, we would have $\lambda = 1$, $z=1$ and $\kappa$ would be dimensionful. We do not put a constant in front of the potential $\mathcal{V}$ because we are going to include an independent coupling constant for each one of its terms.

The potential $\mathcal{V}$ can be, in principle, any FDiff scalar made with the spatial metric $g_{ij}$, the vector
\begin{equation}
 a_i = \frac{ \partial_i N }{N}
\end{equation}
and their FDiff-covariant derivatives (curvature tensors and their derivatives for $g_{ij}$). The potential contains no time derivatives and does not depend on $N_i$. In particular, the $z=1$ potential, which is the most relevant one for the large-distance physics, is 
\begin{equation}
 \mathcal{V}^{(z=1)} = - \beta R - \alpha a_i a^i \,,
\label{z1potential}
\end{equation}
where $\beta$ and $\alpha$ are coupling constants.

The particular formulation of the Ho\v{r}ava theory we study in this paper is related to the behavior of the kinetic term under anisotropic conformal transformations. If the constant $\lambda$ is fixed at the KC point $\lambda = 1/3$, then under the anisotropic conformal transformations
\begin{equation}
 g_{ij} \rightarrow e^{2\Omega} g_{ij} \,,
\hspace{2em}
 N \rightarrow e^{3\Omega} N \,,
\hspace{2em}
 N_i \rightarrow e^{2\Omega} N_i \,,
\hspace{2em}
 \Omega = \Omega(t,\vec{x}) \,,
\label{conformaltrans}
\end{equation}
the kinetic term $\sqrt{g} N ( K_{ij} K^{ij}  - \lambda K^2 )$ remains invariant \cite{Horava:2009uw}. In general the whole theory is not conformally invariant except for the specific case in which the potential itself is conformally invariant under (\ref{conformaltrans}), a situation that we do not consider here. Our interest in bringing the nonprojectable Ho\v{r}ava theory at the KC point comes from the fact that at this point the extra mode is eliminated and the theory acquires the same degrees of freedom of GR \cite{Bellorin:2013zbp}. As we have already commented, this is due to the emerging of two second-class constraints at the KC point. We remark that at the KC point $\lambda = 1/3$ these constraints are always present regardless of the fact that the potential, and hence the full theory, is not conformally invariant.

In the following we present the Hamiltonian formulation of the nonprojectable Ho\v{r}ava theory at the KC point \emph{for a general, unspecified potential} $\mathcal{V}$ \cite{Bellorin:2013zbp}. We denote by $\pi^{ij}$ the momentum conjugated of $g_{ij}$ and by $P_N$ the one of $N$, whereas we regard the shift vector $N_i$ as a Lagrange multiplier. We study the asymptotically flat case, under which the canonical field variables behave asymptotically as
\begin{equation}
g_{ij} - \delta_{ij} = \mathcal{O}(1/r) \,,
\hspace{2em}
\pi^{ij} = \mathcal{O}(1/r^2) \,,
\hspace{2em}
N - 1 = \mathcal{O}(1/r) \,.
\label{asymptoticonditions}
\end{equation}

The only local constraint associated to gauge symmetries that are homotopic to the identity, and hence of first class, is the momentum constraint $\mathcal{H}^i$, 
\begin{equation}
 \mathcal{H}^i \equiv - 2 \nabla_j \pi^{ij} + P_N \partial^i N  = 0 \,,
 \label{momentumconstraint}
\end{equation}
which generates the purely spatial diffeomorphisms. The second-class constraints are
\begin{eqnarray}
 P_N &=& 0 \,,
 \\
 \pi &\equiv& g^{ij} \pi_{ij} = 0 \,,
 \\
  \mathcal{H} &\equiv&
 \frac{2\kappa}{\sqrt{g}} \pi^{ij} \pi_{ij}  
 + \sqrt{g}\, \mathcal{U} = 0 \,,
\label{hamiltonianconstraintgeneral}
 \\
 \mathcal{C} &\equiv&
 \frac{3\kappa}{\sqrt{g}} \pi^{ij} \pi_{ij} 
 - \sqrt{g}\, \mathcal{W} = 0 \,.
 \label{cconstraint}
\end{eqnarray}
$\mathcal{U}$ and $\mathcal{W}$ are derivatives of the potential defined by\footnote{We have modified the original definition of $\mathcal{C}$ given in Ref.~\cite{Bellorin:2013zbp} by dividing it by $N$.}
\begin{eqnarray}
&& \mathcal{U} \equiv
\frac{1}{\sqrt{g}} 
  \frac{\delta}{\delta N} \int d^3y \sqrt{g} N \mathcal{V}
 = \mathcal{V} 
   + \frac{1}{N} \sum\limits_{r=1} (-1)^r 
      \nabla_{i_1 \cdots i_r} \left( N 
         \frac{\partial \mathcal{V}}{\partial ( \nabla_{i_r \cdots i_2} a_{i_1} )} 
          \right) \,,
\nonumber \\
\label{modifiedpotential} \\
&& \mathcal{W} \equiv
  g_{ij} \mathcal{W}^{ij} \,,
\hspace{2em}
{\mathcal{W}}^{ij} \equiv 
  \frac{1}{\sqrt{g} N} 
  \frac{\delta}{\delta g_{ij}} \int d^3y \sqrt{g} N \mathcal{V} \,.
\label{vprima}
\end{eqnarray}
$\nabla_{ij\cdots k}$ stands for $\nabla_{i}\nabla_j \cdots \nabla_{k}$. Adopting the nomenclature of GR, $\mathcal{H}^i = 0$ is called the momentum constraint and $\mathcal{H} = 0$ the Hamiltonian constraint. 

The $\pi = 0$ constraint is the primary constraint that emerges when the theory is formulated at the KC point. Indeed, the conjugated momentum $\pi^{ij}$ obeys the general relation
\begin{equation}
 \frac{\pi^{ij}}{\sqrt{g}} = \frac{1}{2\kappa} G^{ijkl} K_{kl} \,.
\end{equation}
At $\lambda = 1/3$ the hypermatrix $G^{ijkl}$ becomes degenerated, $g_{ij} G^{ijkl} = 0$, which leads directly to the $\pi = 0$ constraint. As a consequence, the secondary constraint $\mathcal{C} = 0$ emerges when the preservation in time of $\pi = 0$ is demanded. Thus, $\pi$ and $\mathcal{C}$ are the two second-class constraints that emerge at the KC point. Unlike GR, in the nonprojectable Ho\v{r}ava theory the Hamiltonian constraint $\mathcal{H}$ is of second-class behavior, which is associated to the fact that it lacks its role as generator of gauge symmetry. Finally, the $P_N = 0$ constraint must be added since in this theory (with $\lambda = 1/3$ or not) we are forced to included the lapse function $N$ as part of the canonical variables.\footnote{An exception for this rule is the model considered in Ref.~\cite{Bellorin:2010je}.}

Unlike GR, the ``bulk" part of the Hamiltonian does not arise as a sum of constraints directly from the Legendre transformation. Instead, it arises in the form
\begin{equation}
 H = \int d^3x \left(
      \frac{2 \kappa N}{\sqrt{g}} \pi^{ij} \pi_{ij} + \sqrt{g} N \mathcal{V} 
      + N_i \mathcal{H}^i \right) \,.
\label{hamiltonianbulk}
\end{equation}
In addition, the boundary term corresponding to the ADM energy \cite{Arnowitt:1962hi}, 
\begin{equation}
E_{\mbox{\tiny ADM}} \equiv \oint d\Sigma_i ( \partial_j g_{ij} - \partial_i g_{jj} ) \,,
\end{equation}
must be incorporated because it is needed for the differentiability of the Hamiltonian under the most general asymptotic variations compatible with asymptotic flatness \cite{Regge:1974zd,Hawking:1995fd}. Specifically, this is a consequence of a contribution of the $z=1$ term $-\beta R$, which asymptotically is of order $\mathcal{O}(1/r^3)$.

By incorporating the constraints $P_N$ and $\pi$, we finally cast the classical Hamiltonian in the form
\begin{equation}
 H = \int d^3x \left(
         \frac{2 \kappa N}{\sqrt{g}} \pi^{ij} \pi_{ij} + \sqrt{g} N \mathcal{V} 
       + N_i \mathcal{H}^i + \sigma P_N + \mu \pi 
         \right)
     + \beta E_{\mbox{\tiny ADM}} \,,
\label{hamiltonianbulkfinal}
\end{equation}
where $N_i$, $\sigma$ and $\mu$ are Lagrange multipliers. This classical Hamiltonian is subject to the constraints (\ref{hamiltonianconstraintgeneral}) and (\ref{cconstraint}), which have not been added with Lagrange multipliers. In Appendix \ref{app:multipliers} we show that if we do so, then the classical condition of preserving the second-class constraints fixes their corresponding Lagrange multipliers equal to zero. Therefore, (\ref{hamiltonianbulkfinal}) is the final classical Hamiltonian and for the classical initial value problem it is enough to impose (\ref{hamiltonianconstraintgeneral}) and (\ref{cconstraint}) only initially (although in the quantum theory there are no such restrictions on the Lagrange multipliers).

The form (\ref{hamiltonianbulkfinal}) of the Hamiltonian is quite suitable for quantization since its bulk part remains nonzero on the constrained phase space. On the other hand, if one wishes to stay as close as possible to GR, then by using the constraint $\mathcal{H} = 0$ this Hamiltonian can also be brought to the form of a sum of constraints in the bulk part plus nontrivial boundary terms. This can be achieved because the difference between $\sqrt{g} N \mathcal{V}$ and $\sqrt{g} N \mathcal{U}$ is a sum of exact divergences, see (\ref{modifiedpotential}), and the only one of these that survives upon integration is the $z=1$ divergence. Thus, we have the identity
\begin{equation}
 \int d^3x \sqrt{g} N \mathcal{U} =
 \int d^3x \sqrt{g} N \mathcal{V} + 2 \alpha \Phi_N \,,
\label{identityuv}
\end{equation}
where
\begin{equation}
 \Phi_N \equiv \oint d\Sigma_i \partial_i N \,.
\label{fluxn}
\end{equation}
The version of the Hamiltonian with a sum-of-constraint bulk part results in\footnote{The presence of the $\Phi_N$ term can also be regarded as a requirement for the differentiability of the Hamiltonian (\ref{hamiltonianfinal}) under general $\delta_N$ variations, since $\mathcal{U}$ has a $2\alpha \nabla_i a^i$ term that asymptotically is of order $\mathcal{O}(1/r^3)$.}
\begin{equation}
H = \int d^3x \left(
N \mathcal{H} + N_i \mathcal{H}^i + \sigma P_N + \mu \pi 
\right) 
+ \beta E_{\mbox{\tiny ADM}} - 2 \alpha \Phi_N \,.
\label{hamiltonianfinal}
\end{equation}
In particular, this form is useful to obtain a simple expression for the energy. It is also useful to address the preservation of all the constraints.

Since the momentum constraint is of first class it is automatically preserved in the totally constrained phase space. In the classical theory, the preservation of the second-class constraints leads to conditions on their associated Lagrange multipliers. In Appendix \ref{app:multipliers} we show that the preservation of $P_N$ and $\pi$ requires the vanishing of the  multipliers of $\mathcal{H}$ and $\mathcal{C}$, as we have already mentioned. Finally, the preservation of $\mathcal{H}$ and $\mathcal{C}$ leads to the following equations for the Lagrange multipliers $\sigma$ and $\mu$:
\begin{eqnarray}
\int d^3y
  \,\sigma\, \frac{\delta}{\delta N} 
	   \int d^3w \sqrt{g}\, \mathcal{U} \delta_{wx}
  + \int d^3y \,\mu\, g_{ij} \frac{\delta}{\delta g_{ij}} 
	   \int d^3w \sqrt{g}\, \mathcal{U} \delta_{wx}
  - \frac{3 \kappa \pi^{ij} \pi_{ij}}{\sqrt{g}} \mu
&& \nonumber \\   
+ 4 \kappa \int d^3y \frac{N \pi_{ij}}{\sqrt{g}} \frac{\delta }{\delta g_{ij}}
       \int d^3w \sqrt{g}\, \mathcal{U} \delta_{wx}
- 4 \kappa \pi^{ij} \mathcal{W}_{ij} = 0 \,, &&
\label{sigmaeq}
\\
  \int d^3y \,\mu\, g_{ij} \frac{\delta}{\delta g_{ij}} 
     \int d^3w \sqrt{g}\, \mathcal{W} \delta_{wx}  
+ \int d^3y \,\sigma\, \frac{\delta}{\delta N} 
     \int d^3w \sqrt{g}\, \mathcal{W} \delta_{wx}
+ \frac{9 \kappa \pi^{ij} \pi_{ij}}{2\sqrt{g}} \mu  
&& \nonumber \\ 
+ 4 \kappa \int d^3y \frac{\pi_{ij}}{\sqrt{g}} \frac{\delta}{\delta g_{ij}} 
     \int d^3w \sqrt{g}\, \mathcal{W} \delta_{wx}
+ 6 \kappa \pi^{ij} \mathcal{W}_{ij}   
= 0 \,. &&
\label{mueq}
\end{eqnarray}
In these expressions we have labeled spatial points with single letters like $w$, $\delta_{wx}$ is the Dirac delta $\delta^{(3)}(w-x)$ and the spatial point $x$ labels the point at which these equations are evaluated. When the potential is of $z=3$ order the analysis of Eqs. (\ref{sigmaeq}) and (\ref{mueq}) shows that they are inhomogeneous \emph{elliptic} partial differential equations of sixth order for $\sigma$ and $\mu$ \cite{Bellorin:2013zbp}.
 
The equations of motion in the Hamiltonian formalism are
\begin{eqnarray} 
\dot{N} &=& N^k \partial_k N + \sigma  \,,
\label{dotn} 
\\
\dot{g}_{ij} &=& 
 \frac{4 \kappa N}{\sqrt{g}} \pi_{ij} + 2 \nabla_{(i} N_{j)}
 + \mu g_{ij} \,,
\label{dotg}
\\
\dot{\pi}^{ij} &=& 
  -\frac{4 \kappa N}{\sqrt{g}} ( \pi^{ik} \pi_k{}^j 
    -\frac{1}{4} g^{ij} \pi^{kl} \pi_{kl} )
  - \sqrt{g} N \mathcal{W}^{ij}
\nonumber \\ & &
  - 2 \nabla_k N^{(i} \pi^{j)k}
  + \nabla_k ( N^k \pi^{ij})
  - \mu \pi^{ij}  \,.
\label{dotpi}
\end{eqnarray}

In the counting of the independent degrees of freedom we have 14 nonreduced canonical variables in the set $\{ ( g_{ij} , \pi^{ij} ) \,,\, ( N,P_N ) \}$, three components of the first-class constraint $\mathcal{H}^i$ and four second-class constraints in the set $\{ P_N, \pi , \mathcal{H} , \mathcal{C} \}$. The number of independent degrees of freedom is given by
\begin{equation}
 (\mbox{14 can. var.}) - \left[ 2 \times (\mbox{3 first-cls. c.}) + ( \mbox{4 second-cls. c.}) \right] = \mbox{4 indep. can. var. }
\end{equation}
Thus, there are two even physical modes in the theory; that is, two modes that propagate themselves with a complete pair of canonical variables. This is the same number of degrees of freedom of GR; there are no extra modes in this theory. This property naturally raises the question whether the dynamics of this theory is able to reproduce the dynamics of GR for suitable large distances, i. e., at least in a perturbative regime for both theories. This was analyzed for the perturbatively linearized theory in Ref.~\cite{Bellorin:2013zbp}; we take again this point in Section \ref{sec:reducedhamiltonian}.

\subsection{Perturbative approach}
\label{sec:perturbations}
In the previous section we summarized the general Hamiltonian formulation applicable to any potential $\mathcal{V}$. In this section we formulate the constraints and the equations for the Lagrange multipliers in an explicit form with the aim of studying rigorously their solutions. Although a complete $z=3$ potential has a huge number of terms, a perturbative approach may render the problem tractable.\footnote{A perturbative study of a $\lambda =1/3$ nonprojectable model without the $a_i$ terms was done in Ref.~\cite{Park:2009hg}. A perturbative analysis of a projectable model was done in Ref.~\cite{Bogdanos:2009uj}.} 
In Ref.~\cite{Colombo:2014lta} Colombo, G\"umr\"uk\c{c}uo\u{g}lu and Sotiriou found that within a $z=3$ potential the nonequivalent terms that contribute to the action quadratic in perturbations (around Minkowski spacetime) are
\begin{eqnarray}
- \mathcal{V}^{(z=1)} &=& 
\beta R + \alpha a_i a^i \,,
\label{v1}
\\
- \mathcal{V}^{(z=2)} &=&
\alpha_1 R \nabla_i a^i 
+ \alpha_2 \nabla_i a_j \nabla^i a^j
+ \beta_1 R_{ij} R^{ij} + \beta_2 R^2 \,,
\label{v2}
\\
- \mathcal{V}^{(z=3)} &=&
\alpha_3 \nabla^2 R \nabla_i a^i 
+ \alpha_4 \nabla^2 a_i \nabla^2 a^i 
+ \beta_3 \nabla_i R_{jk} \nabla^i R^{jk}
+ \beta_4 \nabla_i R \nabla^i R \,,
\nonumber \\
\label{v3}
\end{eqnarray}
where $\nabla^2 \equiv \nabla_i \nabla^i$ and all the alphas and betas are coupling constants.\footnote{In addition to these terms, mixed derivative terms that combine spatial with time derivatives of the spatial metric can be included \cite{Pospelov:2010mp}. They also contribute to the second-order action; actually the main focus of Ref.~\cite{Colombo:2014lta} was on them. These terms could lead to interesting extensions of the Ho\v{r}ava theory. Here we do not consider mixed derivative terms.}

We start the perturbations around Minkowski spacetime by introducing the variables $h_{ij}$, $p_{ij}$ and $n$ in the way
\begin{equation}
 g_{ij} = \delta_{ij} + \epsilon h_{ij} \,,
\hspace{2em}
 \pi^{ij} = \epsilon p_{ij} \,,
\hspace{2em}
 N = 1 + \epsilon n \,.
\label{perturbativevariables}
\end{equation}
We use the orthogonal transverse/longitudinal decomposition
\begin{equation}
 h_{ij} = 
 h_{ij}^{TT} 
 + \frac{1}{2} ( \delta_{ij} - \partial_{ij} \partial^{-2} ) h^T
 + \partial_{(i} h^L_{j)} \,,
\label{decomposition}
\end{equation}
where $\partial_{ij\cdots k}$ stands for $\partial_i\partial_j\cdots\partial_k$, $\partial^2 = \partial_i \partial_i$ and $\partial^{-2} = 1/\partial^2$. $h_{ij}^{TT}$ is subject to $\partial_i h_{ij}^{TT} = h_{ii}^{TT} = 0$. We make an analogous decomposition on $p_{ij}$. We impose the transverse gauge, 
\begin{equation}
 \partial_i h_{ij} = 0 \,,
\label{gauge}
\end{equation}
under which all the longitudinal sector of the metric is eliminated. 

We study the constraints (\ref{momentumconstraint} - \ref{cconstraint}) of the theory at linear order in perturbations adopting the potential defined in (\ref{v1} - \ref{v3}). The momentum constraint (\ref{momentumconstraint}), simplified by using $P_N = 0$ explicitly, eliminates the longitudinal sector of $p_{ij}$,
\begin{equation}
 \partial_i p_{ij} = 0 \,,
\end{equation}
whereas the $\pi = 0$ constraint dictates that $p_{ij}$ is traceless, hence $p^T = 0$. So far we are left with the set $\{ h^{TT}_{ij}, p^{TT}_{ij}, h^T , n\}$ as the set of remaining canonical variables.

Now we move to the $\mathcal{H}$ and $\mathcal{C}$ constraints. To present the results in a compact form, we introduce the vector $\phi$ of scalars and the functional matrix $\mathbb{M}$ in the way
\begin{equation}
 \phi = \left( 
         \begin{array}{c}
                 h^T \\
                 n
         \end{array}\right) \,,
\hspace{2em}
 \mathbb{M} = \left(
              \begin{array}{cc}
                \mathbb{D}_1 & \mathbb{D}_2 \\[1ex]
                \mathbb{D}_2 & \mathbb{D}_3
              \end{array}\right) \,,
\label{phiM}
\end{equation}
where
\begin{equation}
\begin{array}{l}
{\displaystyle
\mathbb{D}_1 \equiv
\frac{1}{8} \left(
( 3 \beta_3 + 8 \beta_4 ) \partial^6 
- ( 3 \beta_1 + 8 \beta_2 ) \partial^4 + \beta \partial^2 
\right) \,, }
\\[2ex]
{\displaystyle
\mathbb{D}_2 \equiv
\frac{1}{2} \left(
\alpha_3 \partial^6 + \alpha_1 \partial^4 + \beta \partial^2 \right)
\,, \hspace{2em}
\mathbb{D}_3 \equiv 
\alpha_4 \partial^6 - \alpha_2 \partial^4 + \alpha \partial^2 } \,.
\end{array}
\label{operators}
\end{equation}
Thus, with the potential given in (\ref{v1} - \ref{v3}), the $\mathcal{H}$ and $\mathcal{C}$ constraints at linear order become
\begin{equation}
 \mathbb{M} \phi = 0 \,,
\label{eqnh}
\end{equation}
where the first row of this vectorial equation represents the $\mathcal{C}$ constraint and the second row the $\mathcal{H}$ constraint. With (\ref{eqnh}) we confirm the consistency of the structure of constraints: (\ref{eqnh}) is a system of sixth-order elliptic partial differential equations for $h^T$ and $n$ (after imposing the appropriated positivity conditions on the matrix of coupling constants).

To solve the constraints (\ref{eqnh}) we start by decoupling them; that is, we want two separate equations in which $h^T$ and $n$ are not mixed. To this end we multiply Eq.~(\ref{eqnh}) with
\begin{equation}
 \left( \begin{array}{rr}
          \mathbb{D}_3  & -\mathbb{D}_2 \\
          -\mathbb{D}_2 & \mathbb{D}_1 
        \end{array} \right)
\end{equation}
from the left and get a diagonal matrix acting on $\phi$, which we write as 
\begin{equation}
 \mathbb{L}\phi = 0 \,,
\hspace{2em}
 \mathbb{L} \equiv  
 \mathbb{D}_1 \mathbb{D}_3 - \mathbb{D}_2^2  \,.
\label{decoupledeq}
\end{equation} 
Equation (\ref{decoupledeq}) represents two decoupled equations for $h^T$ and $n$ and, moreover, the equations are the same (with the same boundary conditions).

Given the values of all the coupling constants, the generic case is when the operator $\mathbb{L}$ is a sixth-order polynomial on $\partial^2$. We can always factorize it; in particular, we may write it as
\begin{equation}
 \mathbb{L} = 
 K ( \partial^2 - z_1 ) P^{(5)}(\partial^2)  \,,
\label{factorizedl}
\end{equation} 
where $P^{(5)}(u)$ is a fifth-order polynomial on $u$, $z_1$ stands for any one of the roots of $\mathbb{L}$, and we first suppose that $K = (1/8) \left(\alpha_4( 3 \beta_3 + 8 \beta_4 ) - 2\alpha_3^2\right) $ is not zero. By combining (\ref{factorizedl}) with (\ref{decoupledeq}) we write the constraints in the form
\begin{equation}
 \partial^2 P^{(5)}(\partial^2) \phi = z_1 P^{(5)} (\partial^2) \phi \,.
\label{eigenfunction}
\end{equation}
The decoupled equation (\ref{eigenfunction}) implies that $P^{(5)}(\partial^2) \phi$ is an eigenfunction of the Laplacian $\partial^2$. Since we are studying the asymptotically flat case, the spatial domain of the problem is the whole $\mathbb{R}^3$ and the boundary condition is that $\phi$ and its derivatives are zero at spatial infinity. Actually, on a noncompact domain, the flat Euclidean Laplacian $\partial^2$ has no nonzero eigenfunctions that go asymptotically to zero in all angular directions. Thus, the only solution of (\ref{eigenfunction}) that satisfies the boundary condition is
\begin{equation}
 P^{(5)}(\partial^2) \phi = 0 
\label{zerosolution}
\end{equation}
everywhere.

Let us present the same argument in another form. Consider the operator $\partial^2 - z_1$, with $z_1 \in \mathbb{C}$, acting on the space of functions $\psi$ whose domain is the whole $\mathbb{R}^3$ and that go asymptotically to zero (see (\ref{asymptoticonditions})). Thus, Eq.~(\ref{eigenfunction}) can be cast as
\begin{equation}
 ( \partial^2 - z_1 ) \psi = 0 \,.
\label{inverseproblem}
\end{equation}
In the space of functions $\psi$, $\partial^2$ has a continuum spectrum valued in $( - \infty , 0 ]$; it has no eigenvalues. With the prescribed asymptotic behavior the inverse $( \partial^2 - z_1 )^{-1}$ exists for any value of $z_1$, but it behaves in different ways depending on whether $z_1$ belongs to the spectrum or not. If $z_1 \not\in ( - \infty , 0 ]$ the inverse $( \partial^2 - z_1 )^{-1}$ is a bounded operator. In this case Eq.~(\ref{inverseproblem}) automatically implies $\psi = 0$, as stated in (\ref{zerosolution}). If $z_1 \in ( - \infty , 0 ]$, $( \partial^2 - z_1 )^{-1}$ still exists but it is an unbounded operator. However, the right-hand side of Eq.~(\ref{inverseproblem}) is zero; $( \partial^2 - z_1 )^{-1}$ acting on it gives zero anyway. Therefore, for any value of $z_1$, Eq.~(\ref{inverseproblem}) has the function $\psi = 0$ as its only solution satisfying the prescribed asymptotic behavior.

Coming back to Eq.~(\ref{zerosolution}), it turns out that it poses another eigenfunction problem for the Laplacian since its left-hand side is another polynomial on $\partial^2$, such that we may factorize it again,
\begin{equation}
 \left( \partial^2 - z_2 \right) P^{(4)}(\partial^2) \phi = 0 \,.
\end{equation}
Since the same arguments hold to solve this equation, we have $P^{(4)} (\partial^2) \phi = 0$ as the unique solution. We may proceed iteratively continuing with this last equation to finally show that the linear-order versions for the variables $h^T$ and $n$ are equal to zero.

We remark that it is the noncompactness of the domain and the prescribed asymptotic conditions of the problem posed in (\ref{eigenfunction}) that force the everywhere-vanishing function to be the unique eigenfunction.

If $\mathbb{L}$ is a lower order polynomial ($K=0$), an analogous eigenfunction problem for the Laplacian arises since we may factorize the given polynomial. By applying the same reasoning of above, we eventually arrive at the same zero solution. Therefore, we conclude that the unique solution of the linearized $\mathcal{H}$ and $\mathcal{C}$ constraints, which are expressed in (\ref{eqnh}), is
\begin{equation}
 h^T = n = 0 \,.
\end{equation}

There remains a condition in the space of parameters: we require that the whole operator $\mathbb{L}$ is not completely zero since otherwise the number of constraints effectively reduces and additional modes appear. In addition, we know that the perturbatively linearized version of the purely $z =1$ theory is equivalent to perturbatively linearized GR \cite{Bellorin:2013zbp}. To combine these two facts, we require that the fourth-order coefficient of $\mathbb{L}$, associated to the $z=1$ operators of the theory, is nonzero,
\begin{equation}
 \beta (2 \beta - \alpha)  \neq 0 \,.
\label{boundz1}
\end{equation}
We regard this as a condition for the continuity in the number of degrees of freedom and for having a weak regime that tends to GR.

The perturbative version of Eqs.~(\ref{sigmaeq}) and (\ref{mueq}) is obtained by regarding the Lagrange multipliers as variables of first order in perturbations. The linearized version of (\ref{sigmaeq} - \ref{mueq}) forms a system equivalent to (\ref{eqnh}),
\begin{equation}
\mathbb{M} \left(
  \begin{array}{c}
      \mu \\
      \sigma
  \end{array} \right)
 = 0
\,.
\label{sigmamueq}
\end{equation}
Thus, by applying the same procedure as above, we obtain that $\sigma$ and $\mu$ are zero at linear order in perturbations.

With all this information we may evaluate directly on Eq.~(\ref{dotg}) the condition of preservation in time of the transverse gauge (\ref{gauge}) (which is a canonical gauge). Considering the perturbation $N_i = \epsilon n_i$, Eq.~(\ref{dotg}) at linear order in perturbations yields
\begin{equation}
 \partial^2 n_i + \partial_i \partial_k n_k = 0 \,.
\end{equation}
This equation, combined with the boundary condition $n_i |_{\infty} = 0$, implies $n_i = 0$. We stress that this restriction and (\ref{sigmamueq}) are requirements of the classical formulation. They do not arise in the quantum theory.

We finally have that, when all the constraints have been solved and the gauge has been fixed at linear order, there remains the pair $\{ h^{TT}_{ij} , p^{TT}_{ij} \}$ as the set of free canonical variables. This confirms rigorously the number of two propagating degrees of freedom that the generic and nonperturbative Hamiltonian analysis anticipated.


\section{Focusing the quantization}

\subsection{The reduced Hamiltonian and its spectrum}
\label{sec:reducedhamiltonian}
Once we know the solutions of all the constraints in the transverse gauge,  we may compute the reduced canonical Hamiltonian of the linearized theory. Since in this theory we have the version (\ref{hamiltonianbulkfinal}) for the Hamiltonian with a nonvanishing bulk part, the reduced Hamiltonian is obtained by simple substitution of the solutions of the constraints \emph{at linear order} into the second-order Hamiltonian density (the boundary term of (\ref{hamiltonianbulkfinal}) cancels itself after the substitution). We have seen that at linear order in the transverse gauge it holds $h_i^L = h^T = n = p_i^L = p^T = p_n = 0$. The substitution of these solutions yields
\begin{equation}
H_{\mbox{\tiny RED}} =
\int d^3x \left( 2 \kappa p^{TT}_{ij} p^{TT}_{ij}
+ \frac{1}{4} h^{TT}_{ij} \mathbb{V} h^{TT}_{ij} \right)
\,,
\label{reducedhamiltonian}
\end{equation}
where
\begin{equation}
\mathbb{V} = 
- \beta \partial^2 - \beta_1 \partial^4 
+ \beta_3 \partial^6   \,.
\label{operatorpotential}
\end{equation}

Alternatively, it is interesting to see how this reduced Hamiltonian can be obtained from the version (\ref{hamiltonianfinal}) of the exact Hamiltonian whose bulk part is a sum of constraints but there remains the boundary terms. In Appendix \ref{app:boundary} we show that this can be effectively achieved in a quite similar fashion to the asymptotically flat reduced Hamiltonian of GR. In particular, this requires considering the solutions of the constraints at second order in perturbations. In that appendix we show that the boundary terms give the correct reduced Hamiltonian despite the fact that this is a theory with higher order derivatives.

There is a further connection between this theory and GR. The largest-distance dynamics of the perturbatively linearized theory can be obtained from the reduced Hamiltonian (\ref{reducedhamiltonian}) by neglecting the higher order derivatives against the lowest order one. By doing so we obtain the effective Hamiltonian for the tensorial modes
\begin{equation}
H^{\mbox{\tiny eff}}_{\mbox{\tiny RED}} =
 \int d^3x \left( 2 \kappa p^{TT}_{ij} p^{TT}_{ij}
   - \frac{\beta}{4} h^{TT}_{ij} \partial^2 h^{TT}_{ij} \right)
\,.
\label{wavehamiltonian}
\end{equation}
This is equivalent to taking only the $z=1$ potential (\ref{z1potential}) and then linearizing it \cite{Bellorin:2013zbp}. Thus, the perturbatively linearized version of the large-distance effective action is physically equivalent to linearized GR. Here one of the key features is the vanishing of the variables $h^T$ and $n$ at linear order in perturbations. The evolution equations arising from (\ref{wavehamiltonian}) constitute the wave equation for $h^{TT}_{ij}$, thus the perturbative large-distance theory around Minkowski spacetime propagates gravitational waves exactly as linearized GR does. However, the nonperturbative dynamics of both theories are different, even considering only the $z=1$ order in the side of the Ho\v{r}ava theory, since the nonperturbative field equations are different.

The requirement of positivity of the reduced Hamiltonian imposes constraints on the coupling constants $\beta$, $\beta_1$ and $\beta_3$ (we assume that $\kappa$ is positive). We require that $\mathbb{V} \geq 0$. Consequently, from the dominant term in the low-energy range we have that $\beta > 0$ and from the one of the high-energy range it follows $\beta_3 < 0$ ($\beta = 0$ is excluded by (\ref{boundz1}) and $\beta_3 = 0$ is excluded in order to have a genuine $z=3$ Hamiltonian). There is also a bound on $\beta_1$, whose all possible values we consider in the following.
\begin{enumerate}

\item Case $\beta_1 \leq 0$. In this case $\mathbb{V} \geq 0$ automatically at all ranges of energy.
	
\item Case $\beta_1 > 0$. We address this case by proposing the factorization of $\mathbb{V}$,
\begin{equation}
 \mathbb{V} = 
 \beta_3 \partial^2 ( \partial^2 - z_+ ) ( \partial^2 - z_- ) \,,
\end{equation}
where 
\begin{equation}
 z_{\pm} =
 \frac{1}{2\beta_3}
 \left( \beta_1 \pm \sqrt{ \beta_1^2 + 4 \beta \beta_3 } \right)\,.
\end{equation}

\renewcommand{\labelenumii}{\theenumi.\arabic{enumii}}	
\begin{enumerate}
		
\item If the discriminant is nonpositive, $\beta_1^2 + 4 \beta \beta_3 \leq 0$ we have that $z_- = \bar{z}_+$. The potential $\mathbb{V}$ is positive, since, for a test function $\psi$, its integral can be written as
\begin{equation}
 \beta_3 \int d^3x \, \bar{\psi}\, \partial^2 ( \partial^2 - \bar{z}_+ ) ( \partial^2 - z_+ ) \psi
 =
 - \beta_3 \int d^3x \, | ( \partial^2 - z_+ ) \partial_i \psi|^2 \,.
\end{equation}

\item If the discriminant is positive, $\beta_1^2 + 4 \beta \beta_3 > 0$, $z_{\pm}$ are real and, due to the signs of the coupling constants, both are negative. The Fourier transform (FT) of $\mathbb{V}$, which is
\begin{equation}
 \tilde{\mathbb{V}}(k^2) = 
 |\beta_3| k^2 ( k^2 - |z_+| ) ( k^2 - |z_-| ) \,,
\label{potentialfourier}
\end{equation}
is useful for determining whether the spectrum of $\mathbb{V}$, given by all the values $\nu$ for which there is no solution $\psi$ of the equation
\begin{equation}
 ( \mathbb{V} - \nu ) \psi = g \,,
\label{eqnu}
\end{equation}
is positive. The function (\ref{potentialfourier}) is a real-valued third-order polynomial of $k^2$. In Fig.~\ref{fig:potential} we show a plot of $\tilde{\mathbb{V}}$ exhibiting its characteristic form in this case. It has a global minimum, which we denote as $\tilde{\mathbb{V}}_0$, and it does not have a global maximum. For our purposes we also need to know that $\tilde{\mathbb{V}}_0$ is always negative, as indicated in the plot.

\begin{figure}[!ht]
	\begin{center}
		\includegraphics[scale=0.35]{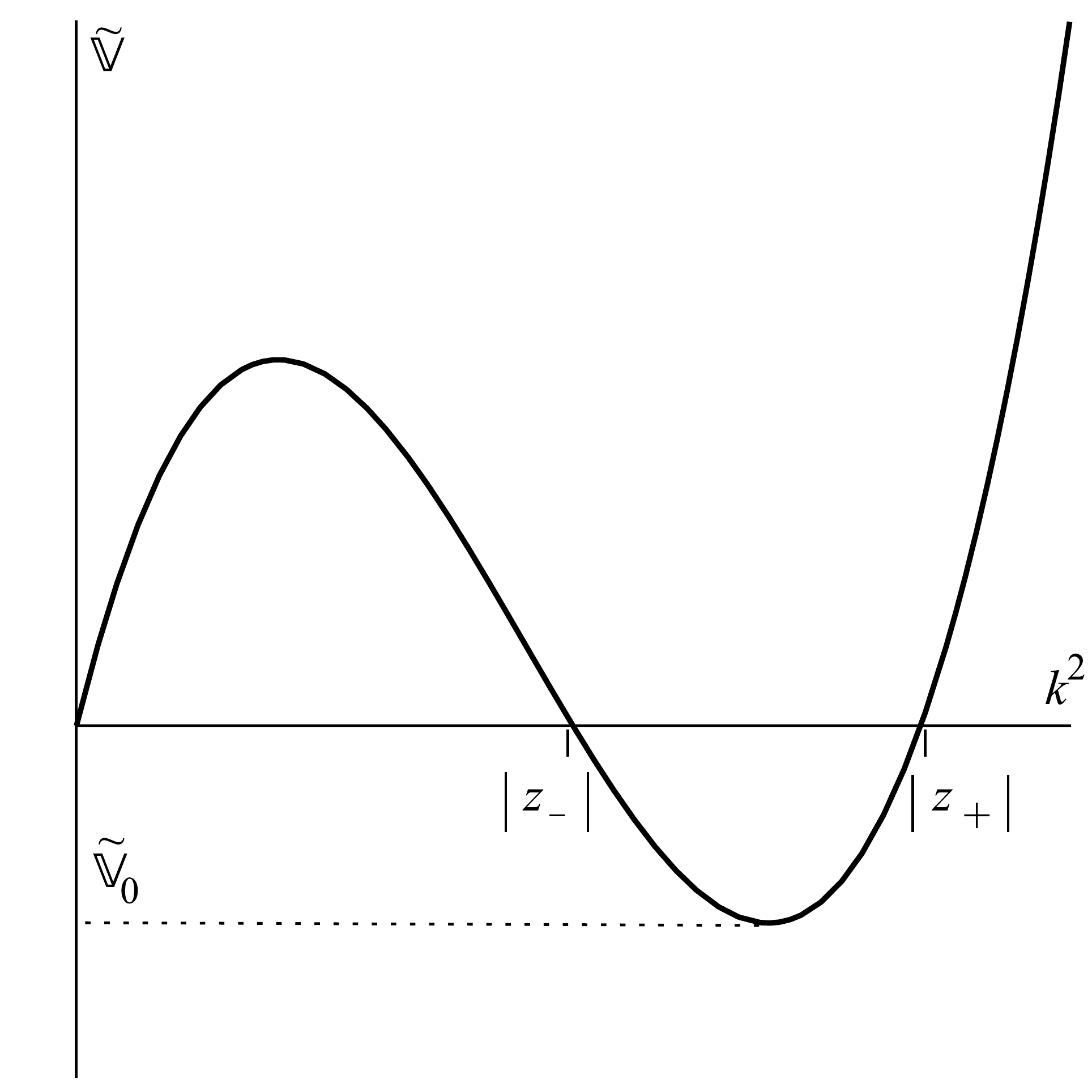}
		\caption{\label{fig:potential} \small The Fourier transform of the operator $\mathbb{V}$ in the case 2.2.}
	\end{center}
\end{figure}

The solutions of (\ref{eqnu}) for all $\nu \in \mathbb{C}$ go as follows: if $\nu$ has a nonzero imaginary part, then the solution of (\ref{eqnu}) exists and its FT is given by
\begin{equation}
 \tilde{\psi} = \frac{\tilde{g}}{\tilde{\mathbb{V}} - \nu} \,.
\label{nusolution}
\end{equation}
If $\nu$ is real and satisfies $\nu < \tilde{\mathbb{V}}_0$ then the solution of (\ref{eqnu}) is also given by (\ref{nusolution}). Finally, if $\nu$ is real and satisfies $\nu \geq \tilde{\mathbb{V}}_0$ then the expression (\ref{nusolution}) has a pole, the solution of (\ref{eqnu}) does not exist. We conclude that in this case the spectrum is formed by all the real values $\nu$ that satisfy $\nu \geq \tilde{\mathbb{V}}_0$. Since $\tilde{\mathbb{V}}_0$ is negative, the spectrum is not positive definite.
		
\end{enumerate}
	
\end{enumerate}
Case 2.1 can be cast as the range in $\beta_1$ given by $0 < \beta_1 \leq 2\sqrt{\beta|\beta_3|}$. Therefore, the union of cases 1 and 2.1, which are the ones with a positive spectrum of $\mathbb{V}$, is $\beta_1 \leq 2\sqrt{\beta|\beta_3|}$.

In summary, the restrictions on the coupling constants needed for the continuity in the number of degrees of freedom, weakest regime approaching to GR and positivity and $z=3$ behavior of the Hamiltonian are
\begin{equation}
 \alpha \neq 2\beta \,,
 \hspace{2em}
 \beta > 0 \,,
 \hspace{2em}
 \beta_3 < 0 \,,
 \hspace{2em}
 \beta_1 \leq 2 \sqrt{\beta |\beta_3|} \,.
\end{equation}


\subsection{The propagator of the physical modes}
Upon the results of the previous sections on the linearized theory, in this section we obtain the propagators of the independent physical modes in the transverse gauge, which for the full $z=3$ KC Ho\v{r}ava theory it has not been considered previously. With the propagator at hand and with the knowledge of the generic structure of the interactions we may compute the superficial degree of divergence of 1PI diagrams and discuss the power-counting renormalizability.

The path integral in terms of the reduced phase space is\footnote{In formulas like (\ref{pathintegralreducedcan}) we omit product symbols like $\prod\limits_{i \leq j}{ \mathcal{D}h^{TT}_{ij} }$, etc.}
\begin{equation}
 Z_0 = 
 \int \mathcal{D}h_{ij}^{TT} \mathcal{D}p^{TT}_{ij} 
 \exp\left[ i\int dt d^3x \left( 
 p^{TT}_{ij} \dot{h}_{ij}^{TT} - \mathcal{H}_{\mbox{\tiny RED}} \right) \right] \,,
\label{pathintegralreducedcan}
\end{equation}
where the reduced Hamiltonian density $\mathcal{H}_{\mbox{\tiny RED}}$ can be read from (\ref{reducedhamiltonian}). After a Gaussian integration in $p^{TT}_{ij}$ we obtain the path integral in the noncanonical form
\begin{equation}
Z_0 = 
\int \mathcal{D}h_{ij}^{TT}
\exp\left[ \frac{i}{4} \int dt d^3x \left( 
 \frac{1}{2\kappa} \dot{h}^{TT}_{ij} \dot{h}^{TT}_{ij} 
 - h^{TT}_{ij} \mathbb{V} h^{TT}_{ij}
 \right) \right] \,.
\label{pathintegralreduced}
\end{equation}
Consequently, the full propagator of the physical modes is 
\begin{equation}
 \left< h_{ij}^{TT} h^{TT}_{kl} \right> =
 \frac{P^{TT}_{ijkl}}{\omega^2 / 2\kappa - \beta \vec{k}^{\,2} 
 	     + \beta_1 \vec{k}^{\,4} + \beta_3 \vec{k}^{\,6}} \,,
\label{propagator}
\end{equation}
where
\begin{equation}
 {\displaystyle 
 P^{TT}_{ijkl} \equiv 
 \frac{1}{\sqrt{2}} \left( 
   \theta_{ik} \theta_{jl} + \theta_{il} \theta_{jk} 
 - \theta_{ij} \theta_{kl} \right) } \,,  
\hspace{2em}
 {\displaystyle \theta_{ij} \equiv \delta_{ij} 
 	- \frac{k_i k_j}{\vec{k}^{\,2}} } \,.
\end{equation}
Notice that only some terms of the potential (\ref{v1} - \ref{v3}) contribute to the propagator of the physical modes. The independent propagator (\ref{propagator}) of this theory behaves just as was the aim in the original formulation of Ho\v{r}ava for having a renormalizable and unitary theory of quantum gravity \cite{Horava:2009uw}: for high $\omega$ and $\vec{p}$ it is dominated by the $z=3$ mode $(\omega^2 / 2 \kappa + \beta_3 \vec{p}^{\:6})^{-1}$ and there are no more independent propagators other than (\ref{propagator}).

With the aim of analyzing UV divergences, we now study qualitatively the structure of the interactions. This requires us to go beyond the linear order. In particular, under the scheme of dealing with reduced variables, the constraints must be solved at higher orders in perturbations. We concentrate ourselves in the second-class constraints since for the first-class one the standard techniques of quantization of gauge systems can, in principle, be applied. 

Among the set of second-class constraints of the theory, $\mathcal{H}$ and $\mathcal{C}$ possess the more involved structure since they are partial differential equations. At higher orders in perturbations their solutions require the inverse of a  nonlocal operator.\footnote{Renormalization of gravity theories with nonlocal terms has been considered in Ref.~\cite{varios:superrenormalizable}, getting super-renormalizable theories.} 
The operator is the matrix $\mathbb{M}$ given in (\ref{phiM}). To illustrate this, we may present the Hamiltonian constraint $\mathcal{H}$ at second order in perturbations, which is
\begin{equation}
\begin{array}{rr} {\displaystyle
 2 \epsilon \left( \mathbb{D}_2 h^T + \mathbb{D}_3 n \right) =
  \frac{\epsilon^2}{4} \left[ 
    - 8 \kappa p^{TT}_{ij} p^{TT}_{ij}
     + \beta_1 \partial^2 h^{TT}_{ij} \partial^2 h^{TT}_{ij} 
     + \beta_3 \partial^2 \partial_i h^{TT}_{jk} 
     \partial^2 \partial_i h^{TT}_{jk} \right.}
\\[2ex] \hspace{2em} {\displaystyle \left.
 + \left( \beta + \alpha_1 \partial^2 + \alpha_3 \partial^4 \right)
   \left( 4 h^{TT}_{ij} \partial^2 h^{TT}_{ij} 
          + 3 \partial_i h^{TT}_{jk} \partial_i h^{TT}_{jk}
          - 2 \partial_i h^{TT}_{jk} \partial_k h^{TT}_{ij} \right) 
 \right] } \,, 
\end{array}
\label{hamiltonianconstraintsecondorder}
\end{equation}
where $\mathbb{D}_2$ and $\mathbb{D}_3$ were defined in (\ref{operators}). In all the terms weighted by a power of $\epsilon^2$ we have substituted the linear-order solutions for the variables that are restricted by the constraints. Note that in the left-hand side member of this constraint we have the second row of the matrix $\mathbb{M}$ acting on the vector $\phi$ (\ref{phiM}). As usual in a perturbative approach, at any order in perturbations the solutions for $h^T$ and $n$ corresponding to the previous orders must be substituted everywhere except on the term of lowest order in $\epsilon$, which is always the one arising in the left-hand member of (\ref{hamiltonianconstraintsecondorder}). Therefore, the $\mathcal{H}$ and $\mathcal{C}$ constraints become linear equations on these variables at any order in perturbations and the operator acting on them is $\mathbb{M}$.

Thus, we see that the solutions of the $\mathcal{H}$ and $\mathcal{C}$ constraints require the use of a nonlocal operator, which in general is difficult to represent. However, for our purposes we only need to know the distribution of momenta at the UV regime. We may then approximate the solutions by taking only the terms that contribute with the highest power of momenta in the Fourier space. To achieve this we make the following observation:  at any order in perturbations, the highest number of spatial derivatives that the $\mathcal{H}$ and $\mathcal{C}$ constraints have \emph{is the same} both for the scalars $h^T$ and $n$ and for the tensorial modes $h_{ij}^{TT}$. This a consequence of two facts: (i) in the decomposition (\ref{decomposition}) $h_{ij}^{TT}$ and $h^T$ enter with the same order in derivatives (or the same power of Fourier-space momentum, if one whishes)\footnote{Some derivatives that act on $h^T$ are missed in $h^{TT}_{ij}$ since it satisfies $\partial_i h^{TT}_{ij} = 0$. However, there remain other combinations that are not divergences on $h^{TT}_{ij}$. In this discussion we are interested only in the powers of momenta, regardless of their origin.}, and (ii) we are considering the presence of all the inequivalent FDiff-covariant interaction terms till order $z=3$, which implies that the highest number of derivatives of the lapse function $N$ is equal to the one of the spatial metric $g_{ij}$. As an example, Eq.~(\ref{hamiltonianconstraintsecondorder}) has a maximum of six derivatives acting on $h^T$, $n$ and $h^{TT}_{ij}$. In addition, the $\mathcal{H}$ and $\mathcal{C}$ constraints have no spatial derivatives of the conjugate momenta. Thus, for second and higher orders in perturbations, the UV-dominant part of the solutions can be modeled in the schematic form
\begin{equation}
h^T \,, n \sim 
 \left( \frac{1}{( \partial_m )^{2z}} 
	( \partial_n )^{2z}  \right) 
   \left( h_{ij}^{TT} \cdots h_{kl}^{TT} \right)\,, 
   \frac{1}{( \partial_m )^{2z}} 
    \left( h^{TT}_{ij} \cdots h^{TT}_{kl} p^{TT}_{pq} p^{TT}_{rs} \right)  \,.
\label{highenergysolution}
\end{equation}
At the highest order in derivatives, the matrix $\mathbb{M}$ can be expressed as the operator $\partial^{2z}$ times a matrix of dimensionless coupling constants, whose determinant is $K = (1/8) \left(\alpha_4( 3 \beta_3 + 8 \beta_4 ) - 2\alpha_3^2\right) $. We assume that $K\neq 0$. We keep the dependence on $p_{ij}^{TT}$ in quadratic form at any order in $\epsilon$ since $\mathcal{H}$ and $\mathcal{C}$ only have quadratic dependence on the exact momentum $\pi^{ij}$. Moreover, solving the constraints $\mathcal{H}^i$ and $\pi$ for $p^L_i$ and $p^T$ does not increase or lower the power in $p^{TT}_{ij}$. In Appendix \ref{app:linearmomentum} we develop this last argument.

In $d+1$ spacetime dimensions the canonically conjugated variable $p_{ij}^{TT}$ scales\footnote{We recall that the assignment of dimensions for coordinates and field variables in Ho\v{r}ava gravity is intentionally made to make the coupling constant $\kappa$ dimensionless \cite{Horava:2009uw}.} with the UV cutoff in momenta $\Lambda$ as $\Lambda^d$. In this theory we intentionally have $z=d$. Then, from the schematic relation (\ref{highenergysolution}) we deduce that the solutions $h^T$ and $n$ do not contribute with powers of momenta in the vertices at any order in perturbations. For example, in a $2z$-order cubic interaction like $h^T h^{TT}_{ij} \partial^6 h_{ij}^{TT}$, after substituting the solution for $h^T$, the vertex still contributes with $2z = 6$ powers of momenta. Therefore, after taking into account the nonlocal nature of the solutions of the second-class constraints, we see that the power counting is not altered by the process of solving them.

Upon these considerations and since we have a genuine $z=3$ propagator we may now discuss the power-counting renormalizability guided by the superficial degree of divergence of general 1PI diagrams over the reduced phase space. For this computation we follow Refs.~\cite{Visser:2009fg,Visser:2009ys}. Further developments on the renormalization of Lorentz-violating theories, in particular, studies on the behavior of the subdivergences, were made in Refs.~\cite{varios:subdivergences}. From the propagator (\ref{propagator}) we deduce that if $\Lambda$ is an UV cutoff for the momenta, then $\Lambda^{z}$ is the cutoff for the energy (up to some constants of proportionality that are irrelevant for our purposes), with $z=3$. Therefore, for each loop in the UV regime we have the contribution
\begin{equation}
 \int d\omega d^dk \rightarrow \Lambda^{d+z} \,,
\end{equation}
while for each propagator
\begin{equation}
 I = \Lambda^{2z} \,.
\end{equation}
In any vertex we can have at most a contribution of $2z$ powers of loop momenta coming from the vertex itself (for vertices that are of $2z$ order in spatial derivatives). If in a 1PI Feynman diagram $L$ is the number of loops, $I$ is the number of internal lines and $V$ is the number of vertices, its superficial degree of divergence $D$ is bounded by
\begin{eqnarray}
 D &\leq& ( d + z ) L + 2 z ( V - I ) 
 \\
   &=& ( d - z ) L + 2 z ( L + V - I ) \,. 
\end{eqnarray}
Now the identity $ L - 1 = I -V $ for graphs is used and in addition in this theory we have $z = d$. Therefore, the superficial degree of divergence is bounded by
\begin{equation}
 D \leq 2z \,.
\end{equation}
This is the bound (8) of Ref.~\cite{Visser:2009ys}, where Lorentz-violating theories with interactions depending on spatial derivatives were considered. This degree of divergence coincides with the highest order operators already included in the bare action (once we extend our potential to include all the $z \leq 3$ terms, not only the operators that contribute to the quadratic action). This leads to the conclusion that the theory is power-counting renormalizable. Unitarity and the criterion of power-counting renormalizability are safe in this theory.
 
\subsection{The path integral in the nonreduced phase space}
\subsubsection{Canonical formulation}
If, unlike the procedure in the previous sections, one wants to avoid the problem of solving the constraints and deals with nonreduced variables, then all of the unsolved constraints must be incorporated into the quantization procedure. At least there are two ways to address the quantization of theories with second-class constraints in nonreduced variables: the Dirac brackets in the operator formalism and the adapted measure in the path-integral formalism \cite{Senjanovic:1976br}. Here we study the path integral.

Let us introduce a common notation for the second-class constraints: $\theta_1 \equiv \pi$, $\theta_2 \equiv P_N$, $\theta_3 \equiv \mathcal{C}$ and $\theta_4 \equiv \mathcal{H}$; and let $\chi^i$ denote a gauge-fixing condition for the freedom of performing spatial diffeomorphisms. The path integral in terms of the nonreduced canonical variables is
\begin{equation}
Z_0 = 
\int \mathcal{D}V \delta(\mathcal{H}^i) \delta(\chi^i) \delta(\theta_m)
e^{ i S_{\mbox{\tiny CAN}} } \,,
\end{equation}
where the measure and the action are given by
\begin{eqnarray}
\mathcal{D}V &\equiv& 
\mathcal{D}g_{ij} \mathcal{D}\pi^{ij} 
\mathcal{D}N \mathcal{D}P_N \times
 \det\{ \mathcal{H}^k , \chi^l \} 
\sqrt{ \det \{ \theta_p , \theta_q \} } \,,
\label{measurecanonicalgeneral}
\\
S_{\mbox{\tiny CAN}} &=& 
\int dt \left[ \int d^3x \left( 
  \pi^{ij} \dot{g}_{ij} + P_N \dot{N} 
  - \frac{2 \kappa N}{\sqrt{g}} \pi^{ij} \pi_{ij} 
  - \sqrt{g} N \mathcal{V} \right) 
+ \beta E_{\mbox{\tiny ADM}} \right] \,.
\nonumber \\
\end{eqnarray}
In the canonical formalism the shift vector $N_i$ is a Lagrange multiplier, hence it does not arise in the path integral (unless one wants to ``raise" the $\delta(\mathcal{H}^i)$ up to the Lagrangian).

There is an important simplification in the matrix of Poisson brackets between the second-class constraints that helps to implement the path integral: all the combinations of brackets between the constraints $P_N$ and $\pi$ vanish. Thus, the matrix of brackets acquires the triangular form
\begin{equation}
 \{ \theta_p , \theta_q \} =
 \left( \begin{array}{cc}
   0 & \mathcal{M} \\
   - \mathcal{M}^t & \mathcal{N}
   \end{array} \right) \,,
\end{equation}
where $\mathcal{M}$ is the submatrix of brackets corresponding to the sector $\{ \theta_{p = 1,2} , \theta_{q = 3,4} \}$ and $\mathcal{N}$ is the submatrix of the sector $\{ \theta_{p = 3,4} , \theta_{q=3,4} \}$.
Consequently, the measure for the second-class constraints simplifies,
\begin{equation}
 \sqrt{ \det \{ \theta_p , \theta_q \} } = 
 \det \mathcal{M}  \,.
\label{determinant}
\end{equation}
On the basis of this relation we can incorporate the measure to the Lagrangian by means of fermionic ghosts. For a potential $\mathcal{V}$ the entries of $\mathcal{M}$ are the equal-time brackets
\begin{eqnarray}
\{ P_N(x) , \mathcal{H}(y) \} &=&
- \frac{\delta}{\delta N(x)} 
\int d^3w \sqrt{g}\, \mathcal{U} \delta_{wy}
\,,
\label{bracketph}
\\
\{ P_N(x) , \mathcal{C}(y) \} &=&
\frac{\delta}{\delta N(x)}
\int d^3w \sqrt{g}\, \mathcal{W} \delta_{wy}
\,,
\\ 
\{ \pi(x) , \mathcal{H}(y) \} &=&
\frac{3 \kappa}{\sqrt{g}}  \pi^{ij} \pi_{ij} \delta_{xy}
- \left( g_{ij} \frac{\delta}{\delta g_{ij}} \right)_{\!\!x}
\int d^3w \sqrt{g}\, \mathcal{U} \delta_{wy}
\,,
\\
\{ \pi(x) , \mathcal{C}(y) \} &=&
\frac{9 \kappa}{2 \sqrt{g}} \pi^{ij} \pi_{ij} \delta_{xy}
+ \left( g_{ij} \frac{\delta}{\delta g_{ij}} \right)_{\!\!x}
\int d^3w \sqrt{g}\, \mathcal{W} \delta_{wy}
\,.
\label{bracketpic}
\end{eqnarray}

The vanishing of the brackets between $P_N$ and $\pi$ suggests that perhaps this theory could be reformulated as a theory without second-class constraints and with enhanced gauge symmetries. This technique consists of promoting $P_N$ and $\pi$ to first-class constraints, $\mathcal{H}$ and $\mathcal{C}$ are regarded as gauge-fixing conditions for the associated gauge symmetries and the Hamiltonian is modified without altering the physics. In Appendix \ref{app:firstclass} we study this possibility for the linearized theory, finding eventually that this procedure simply leads to the reduced theory with a trivial gauge symmetry.

With the aim of getting explicit formulas, we now consider the path integral of the linearized theory. We introduce the perturbative variables according to (\ref{perturbativevariables}) and adding $P_N = \epsilon p_n$. We perform the transverse-longitudinal decomposition (\ref{decomposition}) in $h_{ij}$ and $p_{ij}$. We consider all the constraints up to linear order in $\epsilon$ on the measure and deltas and consider the action up to second order in $\epsilon$.

Some variables that we are not interested in can be quickly eliminated along the same lines of Section \ref{sec:perturbations}. The transverse gauge (\ref{gauge}) and the linearized constraints, except $\mathcal{H}$ and $\mathcal{C}$, yield $h_i^L = p_i^L = p^T = p_n = 0$. Recalling our analysis of the linearized $\mathcal{H}$ and $\mathcal{C}$ constraints of Section \ref{sec:perturbations}, we have that the delta factors in the linearized theory become
\begin{equation}
\delta(\mathcal{H}^i) \delta(\chi^i) \delta(\theta_m) =
\delta(p^L_i) \delta(h^L_i) \delta(p_n) \delta(p^T) 
\delta(\mathbb{M} \phi) \,,
\end{equation}
where $\phi$ and $\mathbb{M}$ were defined in (\ref{phiM}). In the passage to the variables $h_i^L$ and $p_i^L$ the factor $ \det\{ \mathcal{H}^k , \chi^l \} $ of (\ref{measurecanonicalgeneral}) is automatically canceled. Taking advantage of the four first deltas we automatically perform the integration in $p^L_i$, $h^L_i$, $p_n$ and $p^T$. This leaves us with the variables $h^T$ and $n$ as the remaining scalars, keeping in mind that the integration in $p^T$ and $p_n$ has already eliminated their propagation.

Because of linearity, the submatrix $\mathcal{M}$ introduced in (\ref{determinant}) becomes equal to the matrix $\mathbb{M}$ defined in (\ref{phiM}). Thus, for the linearized theory we have
\begin{equation}
 \sqrt{ \det\{ \theta_p , \theta_q \} } =
 \det \mathbb{M} \,.
\label{measurelinear}
\end{equation}

After these steps the path integral of the linearized theory becomes
\begin{equation}
Z_0 = \int \mathcal{D}V \delta(\mathbb{M}\phi)
\exp{ \left[ 
 i \epsilon^2 \int dt d^3x \left( 
	p^{TT}_{ij} \dot{h}^{TT}_{ij} - \mathcal{H}_{\mbox{\tiny RED}} 
	- \phi^t \mathbb{M} \phi \right) 
	   \right]} \,,
\label{pathintegralpreliminar}
\end{equation}
where now
\begin{equation}
 \mathcal{D}V = 
 \mathcal{D}h^{TT}_{ij} \mathcal{D}p^{TT}_{ij} \mathcal{D}\phi
 \times \det\mathbb{M}
\label{measurephi} 
\end{equation}
and $\mathcal{H}_{\mbox{\tiny RED}}$ can be extracted from (\ref{reducedhamiltonian}). There is no time derivative for the scalars $h^T$ and $n$, as we anticipated. This reflects the fact that the only propagating degrees of freedom are the transverse-traceless tensorial modes. We also remark on the determinant role of the measure associated to the second-class constraints: since the combination $\det\mathbb{M} \times \delta(\mathbb{M} \phi)$ is equivalent to $\delta(\phi)$, in (\ref{pathintegralpreliminar}) we can perform directly the integration in $\phi$. The resulting path integral is exactly expressed in terms of the reduced variables with weight $1$ in the measure, as it should be, coinciding with (\ref{pathintegralreducedcan}).

In the linearized theory we may write the measure $\det\mathbb{M}$ in terms of ghosts. To this end we use two ghost fields $c_1,c_2$ and two antighost fields $\bar{c}_1,\bar{c}_2$. Their contribution to the action is
\begin{equation}
 \int dt d^3x \left( \bar{c}_1 \mathbb{D}_1  c_1 + \bar{c}_1 \mathbb{D}_2 c_2 + \bar{c}_2 \mathbb{D}_2 c_1 + \bar{c}_2 \mathbb{D}_3 c_2 \right)
\end{equation}
The operators $\mathbb{D}_{1,2,3}$, which were defined in (\ref{operators}), are third-order polynomials of the flat Laplacian. Thus, these ghosts/antighost acquire propagators with a $z=3$ scaling in the spatial momenta, but they do not get dependence on the frequency when representing the measure.

We have seen that in the linearized theory the part of the measure corresponding to the second-class constraints is the factor $\det \mathbb{M}$, which has no consequence on the dynamics because it is independent of the fields. However, at higher order in perturbations (or in the nonperturbative theory) the measure $\sqrt{\det\{\theta_p , \theta_q\}}$ depends in a highly nontrivial way on the fields, as can be deduced from (\ref{bracketph} - \ref{bracketpic}). Thus, the second-class constraints together with their associated measure must be carefully considered.

\subsubsection{Recovering the quantum FDiff-covariant action}
In this section we perform an important check of consistency of the quantization procedure: we ask ourselves whether the canonical path integral of the previous section reproduces the action in FDiff-covariant variables and simultaneously we find the appropriated measure for this formalism. To this end it is convenient to avoid the delta in $\phi$ that the canonical path integral (\ref{pathintegralpreliminar}) has since we want to keep the scalars $h^T$ and $n$ as nonzero variables inside the FDiff-covariant action.

By introducing a linear-order Lagrange multiplier $\epsilon b$, where $b$ is a two-component vector of scalars, the delta $\delta (\mathbb{M}\phi)$ in (\ref{pathintegralpreliminar}) can be ``raised up" to the Lagrangian,
\begin{equation}
Z_0 = \int \mathcal{D}V \mathcal{D}b
\exp{ \left( i \epsilon^2 \int dt d^3x \left( 
	p^{TT}_{ij} \dot{h}^{TT}_{ij} - \mathcal{H}_{\mbox{\tiny RED}} 
	- ( \phi - b )^t \mathbb{M} \phi \right ) \right) } \,.
\label{pathintegralpreliminar2}
\end{equation}
By virtue of the self-adjointness of $\mathbb{M}$, the following identity holds:
\begin{equation}
\int d^3x ( \phi - b )^t \mathbb{M} \phi =
\int d^3x \left(
( \phi - \frac{1}{2} b )^t \mathbb{M} (\phi - \frac{1}{2} b )
- \frac{1}{4} b^t \mathbb{M} b \right) \,.
\label{identityM}
\end{equation}
Thus, in the path integral we may perform the following change of variables
\begin{equation}
\phi \rightarrow \phi - \frac{1}{2} b \,,
\end{equation}
which has unit Jacobian. After this change $\phi$ and $b$ are not mixed in the action. The only dependence the resulting action has in $b$ is in the last term of (\ref{identityM}). Since $b$ is a real  bosonic field the integration over it yields a factor $\left(\sqrt{\det\mathbb{M}}\right)^{-1}$ in the measure. Therefore, we have that the path integral with nonzero $h^T$ and $n$ fields take the form
\begin{equation}
Z_0 = 
 \int \mathcal{D}h^{TT}_{ij} \mathcal{D}p^{TT}_{ij}   \mathcal{D}\phi
 \sqrt{ \det \mathbb{M} } 
 \exp{ \left( i \epsilon^2 \int dt d^3x \left( 
	p^{TT}_{ij} \dot{h}^{TT}_{ij} - \mathcal{H}_{\mbox{\tiny RED}} 
	- \phi^t \mathbb{M} \phi \right ) \right) } \,.
\label{finalpathintegralcanonical}
\end{equation}
By contrasting this version with (\ref{pathintegralpreliminar}) we see that the change consists in dropping the delta in $\phi$ at the price of changing the measure. This version of the canonical path integral is also consistent with the formulation in the reduced phase space since the integration over $\phi$ can be directly performed in (\ref{finalpathintegralcanonical}) yielding a factor of $(\sqrt{\det\mathbb{M}})^{-1}$ that cancels itself with the measure.

We now compare with the action written in noncanonical variables (the FDiff-covariant variables). Although those variables give a complete covariant formulation, for simplicity we do the comparison in the transverse gauge, under which (\ref{finalpathintegralcanonical}) is written. The support for this simplification is the fact that the gauge symmetry of pure spatial diffeomorphisms is present in both the Lagrangian and the canonical formulations. The FDiff-covariant variables are the ADM variables $g_{ij}$, $N$ and $N_i$ and the action is given in (\ref{lagrangianaction}). The ghosts associated to the gauge fixing should be included, but they decouple in the linearized theory, thus we do not consider them in this analysis. We introduce the perturbative variables according to (\ref{perturbativevariables}) and adding
\begin{equation}
 N_i = \epsilon ( u_i + \partial_i B ) \,,
\end{equation}
with $\partial_i u_i = 0$.

The linearized version of the action (\ref{lagrangianaction}) in the transverse gauge is given by
\begin{equation}
\begin{array}{rcl}
 S &=&  {\displaystyle
  \epsilon^2 \int dt d^3x \left( 
  \frac{1}{8 \kappa} \dot{h}^{TT}_{ij} \dot{h}^{TT}_{ij} 
+ \frac{1 - 2\lambda}{16 \kappa} (\dot{h}^T)^2
+ \frac{\lambda}{2 \kappa} \dot{h}^T \partial^2 B
 \right.  }
\\ & & {\displaystyle \left.
+ \frac{1 - \lambda}{2\kappa} (\partial^2 B)^2
- \frac{1}{4 \kappa} u_i \partial^2 u_i
- \frac{1}{4} h^{TT}_{ij} \mathbb{V} h^{TT}_{ij}
- \phi^t \mathbb{M} \phi \right) } \,,
\end{array}
\end{equation}
where $\mathbb{V}$ is defined in (\ref{operatorpotential}). To arrive at these expressions we have integrated $h_i^L$ out. According to (\ref{finalpathintegralcanonical}), in the measure of the path integral one must include the factor $\sqrt{\det\mathbb{M}}$. Next, integration in $u_i$ can be performed yielding an irrelevant factor in the denominator of the path-integral integrand. $B$ can also be easily integrated after completing squares, which yields the action
\begin{equation}
 S =  
  \epsilon^2 \int dt d^3x \left( 
  \frac{1}{8 \kappa}\dot{h}^{TT}_{ij} \dot{h}^{TT}_{ij} 
+ \frac{1 - 3\lambda}{16 \kappa ( 1 - \lambda )} (\dot{h}^T)^2
- \frac{1}{4} h^{TT}_{ij} \mathbb{V} h^{TT}_{ij}
- \phi^t \mathbb{M} \phi \right)  \,.
\label{finalcovariantaction}
\end{equation}
The crucial fact about the propagating degrees of freedom at the KC point in the scenario of nonreduced, FDiff-covariant variables can be seen in this action. Recalling that in this theory $\lambda = 1/3$, we have that the action loses the time derivative of $h^T$, whereas the one of $n$ is absent from the very beginning. The goal we pursue in this section is achieved once we compare (\ref{finalcovariantaction}) with (\ref{finalpathintegralcanonical}): with $\lambda = 1/3$ the canonical path integral reproduces the FDiff-covariant Lagrangian since the Gaussian integration of (\ref{finalpathintegralcanonical}) over the momenta $p_{ij}^{TT}$ yields the action (\ref{finalcovariantaction}). With this procedure we have learned that the factor $\sqrt{\det\mathbb{M}}$ must be included in the measure of the path integral in the FDiff-covariant formulation (this factor is not equal to the measure of the second-class constraints in canonical variables!). Again, it is at the level of higher orders in perturbations where this factor affects the dynamics.

\section{The non-kinetic-conformal theory}
\label{sec:nokc}
Since the nonprojectable Ho\v{r}ava theory with $\lambda \neq 1/3$ also has second-class constraints,  in this section we want to consider it briefly with the aim of highlighting the need of incorporating the measure of these constraints to the path integral, as in the case of the KC theory.

The action is of the same form as (\ref{lagrangianaction}), but now with $\lambda \neq 1/3$ (and $\lambda$ otherwise arbitrary, except for requirements of stability of the linearized theory), such that the metric $G^{ijkl}$ has the inverse given by
\begin{equation}
 \mathcal{G}_{ijkl} = 
 \frac{1}{2} (g_{ik} g_{jl} + g_{il} g_{jk} )
 - \frac{\lambda}{3\lambda - 1} g_{ij} g_{kl} \,.
 \label{inverseg}
\end{equation}
For our purposes it is enough to take the large-distance effective action, which has the second-order potential
\begin{equation}
 \mathcal{V} = - \beta R - \alpha\, a_i a^i \,.
\end{equation}

The theory shares with the KC theory the fact that the momentum constraint $\mathcal{H}^i$ is the only first-class constraint. On the other hand, the only second-class constraints are $P_N = 0$ and the Hamiltonian constraint 
\begin{equation}
 \mathcal{H} \equiv
    \frac{2\kappa}{\sqrt{g}} \mathcal{G}_{ijkl}  \pi^{ij} \pi^{kl} 
  + \sqrt{g}\, \mathcal{U} 
  = 0\,,
\label{hamiltonianconstraint}
\end{equation}
where
\begin{equation}
 \mathcal{U} \equiv
  \frac{1}{\sqrt{g}} 
  \frac{\delta}{\delta N} \int d^3y \sqrt{g} N \mathcal{V} =
 - \beta R  +  \alpha ( 2 \nabla_i a^i  + a_i a^i )  \,.
\end{equation}
The Hamiltonian in the nonzero-bulk version takes the form
\begin{equation}
H  =
 \int d^3x \left( \frac{2 \kappa N}{\sqrt{g}} \mathcal{G}_{ijkl} \pi^{ij} \pi^{kl} 
 - \sqrt{g} N ( \beta R + \alpha\, a_i a^i ) + N_i \mathcal{H}^i 
	+ \sigma P_N \right)   \,.
\label{prehamiltonian}
\end{equation}
The preservation in time of $\mathcal{H} = 0$ yields a second-order, linear, elliptic partial differential equation for $\sigma$. With this step the Dirac procedure for analyzing the structure of constraints closes. Since the theory possesses the momentum constraint $\mathcal{H}^i$ as the first-class constraint and the constraints $P_N$ and $\mathcal{H}$ as the second-class ones it results that the theory propagates three even physical modes. Two of them correspond to the two tensorial modes that are also propagated in the KC theory and GR and the other one is the extra scalar mode.

Thus, we have that in this theory there are fewer second-class constraints than in the KC theory. However, as happened in the KC theory, the matrix of Poisson brackets acquires a triangular form since the constraint $P_N$ has a vanishing bracket with itself. Then the measure for the second-class constraints takes the form
\begin{equation}
 \sqrt{\det \{ \theta_p , \theta_q \} } =
 \det \{ P_N , \mathcal{H} \}  \,.
\end{equation}
It can be directly elevated to the Lagrangian by means of fermionic ghosts. The Poisson bracket we need for the measure (evaluated on the constrained phase space) is
\begin{equation}
\{ P_N(x)  , \mathcal{H}(y) \} =
2 \alpha \frac{\sqrt{g}}{N} \left( 
 \nabla_i (\delta_{xy} a^i) - \nabla^2 \delta_{xy} \right) \,.
\label{brackethphi}
\end{equation}
The lesson we extract from this discussion is the fact that also in the nonprojectable Ho\v{r}ava theory with $\lambda \neq 1/3$ the measure of the second-class constraints is needed (as well as the first-class sector), and that it has a nontrivial dependence on the fields whenever one goes beyond the linearized level, which is of course necessary for evaluating interactions. Notice also that, for simplicity, we have restricted ourselves to the large-distance effective action. The measure gets more involved once high-order operators are considered.

\section{Discussion and conclusions}
The nonprojectable Ho\v{r}ava theory \cite{Horava:2009uw,Blas:2009qj} possesses second-class constraints. When it is formulated at the kinetic-conformal point, $\lambda = 1/3$, there are four of them, which, together with the momentum constraint, leave two propagating degrees of freedom. The presence of second-class constraints must be carefully considered in any quantization procedure, since standard techniques for gauge theories that have no second-class constraints could not apply.

One route to deal with the second-class constraints is to solve them. In this direction we have analyzed the perturbative linearized theory in the transverse gauge, taking all the $z=1,2,3$ terms that contribute to the quadratic action. We have found the propagator for the two transverse-traceless tensorial modes. Our perturbative approach confirms that there are no extra modes or ghosts. Moreover, the physical propagator at the UV regime effectively has the scaling in momenta for which the theory was designed. From this and from the qualitative analysis of the vertices we have shown the power-counting renormalizability of the theory. In addition, within the linearized approach we have rigorously corroborated the consistency of the Hamiltonian formulation of the classical theory. We have confirmed that all the differential-equation constraints and conditions for the Lagrange multipliers have elliptic structures and can be consistently solved. We have found conditions on the space of coupling constants needed to ensure the positiveness of the spectrum of the physical Hamiltonian.

To get more insight on the renormalizability of the theory it would be interesting to study the extension of the analysis of Anselmi and Halat, who considered the behavior of subdivergences on Lorentz-violating scalar and fermionic field theories \cite{varios:subdivergences}, to this theory. Those authors found the interesting result that subdivergences in Lorentz-violating theories can be canceled in a similar way as the relativistic theories.

There can be other ways of solving the constraints that could apply even for the nonperturbative theory. These techniques are typically noncovariant (under general spatial transformations). For example, in general relativity this has been broadly undertaken with the light-front coordinates \cite{varios:lightfront}. This approach introduces nonlocal operators in the Lagrangian as a consequence of solving the constraints. The light-front quantization  of quantum chromodynamics uses similar ideas related to null coordinates, see for example \cite{Brodsky:1997de,Srivastava:2000cf}. This has also been applied to electroweak theory \cite{Srivastava:2002mw}. Under this approach the quantization of nonperturbative and perturbative QCD has been focused, even the one-loop renormalization has been obtained \cite{Srivastava:2000cf}. Thus, it would be interesting to explore the possibility of solving the second-class constraints of the nonprojectable Ho\v{r}ava theory using a special coordinate system.

The other route to deal with second-class constraints, which is largely more popular for gauge theories, is to work in the nonreduced phase space. In gauge theories without second-class constraints the standard techniques (Faddeev-Popov and Becchi-Rouet-Stora-Tyutin procedures) have allowed a great advance in establishing their renormalizability (whenever they are so). This has been applied even for general relativity with higher curvature terms \cite{Stelle:1976gc}. However, the point with second-class constraints is that they have no associated gauge symmetry (we have even considered the transformation to a gauge system, but with trivial results).

To start from first principles, we have analyzed the formulation of the path integral with the second-class constraints. We have evaluated the prescription for the measure in the canonical theory, finding that there is a simplification since the square root disappears. We have also found the measure for the nonreduced linearized theory, which confirmed the correctness of the prescribed measure since it leads directly to the reduced canonical theory with measure $1$. The measure can, in principle, be incorporated to the Lagrangian with ghosts, but the propagation of them must be considered carefully since this kind of ghost is not directly connected to gauge symmetries. Indeed, we have seen that they arise with a $z=3$ UV scaling in momenta directly from the measure, but without dependence on the frequency. It would be interesting to explore if at higher orders in perturbations, where the dependence of the constraints on the canonically conjugate momenta (and hence on time derivatives) is activated, one can obtain more information about the dependence on the frequency of the propagation of these ghosts. In general, extracting the consequences the measure associated to second-class constraints has in the dynamics of a given theory is a delicate issue.\footnote{There are exceptions to this rule, for example, the massive Yang-Mills theory, whose measure is dynamically trivial (in the exact theory), such that one can ignore it \cite{Senjanovic:1976br}.}

In the nonreduced scheme we have also applied an approach to reproduce the path integral in terms of FDiff-covariant variables (simply, the ``Lagrangian" approach); in the linearized theory in this case. This procedure yielded the appropriated measure for the Lagrangian formalism. This is a rather nontrivial issue, since if one starts with the pure Lagrangian formulation of the path integral in a theory with second-class constraints, then one has no general recipe for the measure.

Throughout this paper we have used the transverse gauge due to the great simplifications in computations it provides. However, other gauge-fixing conditions can be more convenient for establishing renormalization or for other quantum features. For example, the authors of \cite{Barvinsky:2015kil} found that with a nonlocal gauge-fixing condition they could show the renormalizability of the projectable Ho\v{r}ava theory. The essence of their approach is that with the nonlocal gauge condition they could arrive at regular propagators for all the relevant (nonreduced) variables.


\section*{Acknowledgments}
A. R. is partially supported by Grant Fondecyt No.~1161192, Republic of Chile.

\appendix
\section{The full set of Lagrange multipliers}
\label{app:multipliers}

Here we consider the incorporation of all the secondary constraints to the Hamiltonian. We may start with adding the $\mathcal{H}$ and $\mathcal{C}$ constraints in the form $\int d^3x ( A \mathcal{H} - B \mathcal{C} )$ to the Hamiltonian (\ref{hamiltonianfinal}), where $A$ and $B$ are Lagrange multipliers (signs are for convenience). We obtain the Hamiltonian in the form
\begin{equation}
 H = \int d^3x \left(
       \left( N + A \right) \mathcal{H} + N_i \mathcal{H}^i + \sigma P_N + \mu \pi - B \mathcal{C}  \right) 
     + \beta E_{\mbox{\tiny ADM}} - 2 \alpha \Phi_N \,.
\label{app:hamiltonianprimary}
\end{equation}
We assume that $A$ and $B$ go asymptotically to zero fast enough such that the differentiability of the $z=1$ terms of the Hamiltonian is ensured. Once all the constraints have been incorporated to the Hamiltonian with Lagrange multipliers, the first-class constraint is automatically preserved (weakly vanishing bracket with the Hamiltonian), whereas the preservation of the second-class constraints leads to conditions on the Lagrange multiplier associated to them ($\sigma$, $\mu$, $A$ and $B$). The expression of the $N_i$ multiplier is associated to the chosen gauge-fixing condition.

Preservation of the $P_N = 0$ and $\pi = 0$ constraints yields the following equations for the Lagrange multipliers $A$ and $B$:
\begin{eqnarray}
  \frac{\delta}{\delta N} 
  \int d^3y \sqrt{g} ( A \mathcal{U} + B \mathcal{W} )  &=& 0 \,,
\label{abequationsfulla}
\\
  g_{ij} \frac{\delta}{\delta g_{ij}} 
  \int d^3y \sqrt{g} ( A \mathcal{U} + B \mathcal{W} )
 - \frac{3 \kappa}{\sqrt{g}} \pi^{ij} \pi_{ij} 
    \left( A - \frac{3}{2} B \right)  &=& 0 \,.
\label{abequationsfullb}
\end{eqnarray}
Although these are very involved equations, we can perform a qualitative analysis of their forms, since the structure of the highest derivative terms of Eqs.~(\ref{abequationsfulla} - \ref{abequationsfullb}) can be deduced from inspection. The considerations we make are similar to those done in Ref.~\cite{Bellorin:2013zbp} to conclude that the differential equations for the other Lagrange multipliers, $\sigma$ and $\mu$, are elliptic equations (which we have explicitly checked in the current paper in Section \ref{sec:perturbations}). The main point is that (\ref{abequationsfulla}) and the first term of (\ref{abequationsfullb}) contain second-order functional derivatives of the potential $\mathcal{V}$. The several terms of the potential behave in two ways under these derivatives: there are terms that combine all of their spatial derivative on their coefficients, which are either $A$ or $B$, and terms that yield lower order spatial derivatives on $A$ and $B$. Let us illustrate this with some examples. Employing a nonrigorous but schematic notation, the two $z=3$ terms
\begin{equation}
 \frac{\delta^2}{\delta g_{ij}^2} 
   \left[ \sqrt{g} B \left( \nabla_i R_{jk} \right)^2 \right] \,,
\hspace{2em}
 \frac{\delta^2}{\delta N^2} 
   \left[ \sqrt{g} A \nabla_k a^k \nabla^2 \nabla_l a^l \right]
\label{sixorderterms}
\end{equation}
yield the cubic Laplacian $\nabla^6$ acting on $B$ and $A$ respectively. On the other hand, a term like
\begin{equation}
 \frac{\delta^2}{\delta N^2} 
   \left[ \sqrt{g} A \left( a_k a^k \right)^3 \right]
\end{equation}
does not yield a sixth-order derivative on $A$, but a lower order one. Despite this, we have that Eqs.~(\ref{abequationsfulla} - \ref{abequationsfullb}) yield the operator $\nabla^6$ acting on $A$ and $B$ as their highest order operator because terms like (\ref{sixorderterms}) must be included in the potential either directly or by other terms that give them after integration by parts or using curvature identities. Therefore, we conclude that for a general $z=3$ potential Eqs.~(\ref{abequationsfulla} - \ref{abequationsfullb}) are elliptic equations for $A$ and $B$ (once a condition of positivity of the matrix of coupling constants is imposed). The second crucial property is that Eqs.~(\ref{abequationsfulla} - \ref{abequationsfullb}) form a homogeneous system for $A$ and $B$, unlike the system (\ref{sigmaeq} - \ref{mueq}) for $\sigma$ and $\mu$ that is inhomogeneous. Third, we have the boundary conditions $A,B|_\infty = 0$, thus we expect no other solution than $A = B = 0$.

Let us see how this is verified explicitly in the linearized theory with its general potential. For the linearized theory with the potential (\ref{v1} - \ref{v3}), Eqs. (\ref{abequationsfulla} - \ref{abequationsfullb}) take the form
\begin{equation}
\mathbb{M} \left( \begin{array}{c}
                    B \\
                    A
                  \end{array} \right)
              = 0
\end{equation}
where $\mathbb{M}$ is defined in (\ref{phiM}). Thus, we effectively get a system of sixth-order elliptic equations for $A$ and $B$ (imposing the necessary conditions of signs in the coupling constants). This is the same system of equations we studied in Section \ref{sec:perturbations} for $h^T$ and $n$, with the same boundary condition $A,B|_\infty = 0$. Thus, we have that $A = B = 0$.

\section{The reduced Hamiltonian from boundary terms}
\label{app:boundary}

Similarly to the asymptotically flat case of GR, the reduced Hamiltonian (\ref{reducedhamiltonian}) of the linearized theory can be obtained from the version (\ref{hamiltonianfinal}) of the exact Hamiltonian if one inserts the solution of the constraints \emph{at second order in perturbations} into the boundary terms. From Eqs.~(\ref{fluxn} - \ref{hamiltonianfinal}) we have that these boundary terms, evaluated at second order in perturbations and in the transverse gauge, yield
\begin{equation}
H = - \int d\Sigma_i \left( \beta \partial_i h^T + 2 \alpha \partial_i n \right) 
= - \int d^3x \left( \beta \partial^2 h^T + 2 \alpha \partial^2 n \right) \,.
\label{boundary}
\end{equation}
Although the linear-order solutions for the variables $h^T$ and $n$ are everywhere vanishing, their second-order versions do not. Since they are involved in the energy, we see that the role of the pair $\{h^T,n\}$ at second order is analogous to the role the second-order variable $h^T$ has in linearized GR \cite{Arnowitt:1962hi,Regge:1974zd}. 

There is a simplification in the evaluation of (\ref{boundary}): the combination $\beta \partial^2 h^T + 2 \alpha \partial^2 n$ arises directly inside the second-order Hamiltonian constraint $\mathcal{H}$. Indeed, the constraint $\mathcal{H}$ at second order is written in (\ref{hamiltonianconstraintsecondorder}); here we expand its left-hand side,
\begin{equation}
\begin{array}{ll}
 \beta \partial^2 h^T + 2 \alpha \partial^2 n 
 + \left( \alpha_1 \partial^4 + \alpha_3 \partial^6 \right) h^T
 - 2 \left( \alpha_2 \partial^4 - \alpha_4 \partial^6 \right) n	
	=

\\[1ex] \hspace{2em} {\displaystyle
	\frac{\epsilon}{4} \left[ 
	- 8 \kappa p^{TT}_{ij} p^{TT}_{ij} 
	+ \beta_1 \partial^2 h^{TT}_{ij} \partial^2 h^{TT}_{ij} 
	+ \beta_3 \partial^2 \partial_i h^{TT}_{jk} 
	\partial^2 \partial_i h^{TT}_{jk}
 \right.}
\\[2ex] \hspace{2em} {\displaystyle
	\left.
	+ \left( \beta + \alpha_1 \partial^2 + \alpha_3 \partial^4 \right)
      \left( 4 h^{TT}_{ij} \partial^2 h^{TT}_{ij} 
	+ 3 \partial_i h^{TT}_{jk} \partial_i h^{TT}_{jk}
	- 2 \partial_i h^{TT}_{jk} \partial_k h^{TT}_{ij} \right)
	 \right] } \,.
\end{array}
\end{equation}
In the left-hand side of this constraint there are other terms that depend on $h^T$ and $n$, but they are all exact divergences of higher (fourth and sixth) order that vanish upon volume integration. There are other divergences in the right-hand side that vanish upon integration and also one term cancels itself after the integration due to the transverse gauge. 

Thus, we can solve the second-order combination $\beta \partial^2 h^T + 2 \alpha \partial^2 n$ in terms of $\{h^{TT}_{ij}$, $p^{TT}_{ij}\}$ directly from the $\mathcal{H} = 0$ constraint, with no need of using any other constraint. This is related to the fact that the boundary terms of the Hamiltonian (\ref{hamiltonianfinal}) are needed specifically for the differentiability of the $z=1$ terms of $\int d^3x N \mathcal{H}$. The solution of the second-order Hamiltonian constraint $\mathcal{H}$ is
\begin{equation}
 \int d^3x \left( \beta \partial^2 h^T + 2 \alpha \partial^2 n \right) =
 - \int d^3x \left(  
   2 \kappa p^{TT}_{ij} p^{TT}_{ij}
 + \frac{1}{4} h^{TT}_{ij} \mathbb{V} h^{TT}_{ij} \right) \,.
\end{equation}
Therefore, the reduced Hamiltonian coincides with (\ref{reducedhamiltonian}).

\section{The $\mathcal{H}^i$ and $\pi$ constraints at higher orders}
\label{app:linearmomentum}

Our interest in this appendix is to show that the solutions of the $\mathcal{H}^i$ and $\pi$ constraints for $p^L_i$ and $p^T$ at higher orders in perturbations are always linear in the transverse-traceless component $p^{TT}_{ij}$ and that they are of zero order in momentum in the Fourier space. These results are direct consequences of the facts that $\mathcal{H}^i = 0$ and $\pi =0$ are linear in the conjugate momentum $\pi^{ij}$ and that $p_{ij}^{TT}$, $p^T$ and $\partial_i p^L_j$ are of the same weight in Fourier momentum in the decomposition of $p_{ij}$. 

Let us start with solving the momentum constraint $\mathcal{H}^i$ for the longitudinal component $p^L_i$. The covariant divergence of $\pi^{ij}$ has the expression
\begin{equation}
 \nabla_i \pi^{ij} = \partial_i \pi^{ij} 
  + \Gamma_{ik}^i \pi^{kj} + \Gamma_{ik}^j \pi^{ik} + \Gamma_{ik}^k \pi^{ij} \,.
\label{nablapi}
\end{equation}
We make the perturbation (\ref{perturbativevariables}) together with the decomposition (\ref{decomposition}) in the momentum constraint $\mathcal{H}^i$. The first term in the right-hand side of (\ref{nablapi}) is always of linear order in $\epsilon$. This is the term used to solve for $p^L_i$ since the terms $\Gamma \times \pi$ are of quadratic order and higher in $\epsilon$. Thus, at any order in $\epsilon$ the $\mathcal{H}^i$ constraint can be solved in the following way 
\begin{equation}
\epsilon \left( \delta_{ij} \partial^2 + \partial_{ij} \right) p^L_i =
\epsilon \left.\left( 
  \Gamma_{ik}^i p_{kj} + \Gamma_{ik}^j p_{ik} + \Gamma_{ik}^k p_{ij} \right)\right|_{\mbox{\tiny lower order sol.}} \,.
\label{eqpl}
\end{equation}
We must take into account that the minimum order in $\epsilon$ of $\Gamma_{ij}^k$ is one. The solution for $\partial_i p^L_j$ in terms of $p_{ij}^{TT}$ does not contribute with powers of momenta in Fourier space since there arises the inverse of a linear-derivative operator multiplied by a factor of $\Gamma_{kl}^m$. Schematically,
\begin{equation}
 \partial_i p^L_j \sim \frac{1}{\partial_k} \left(\Gamma_{lm}^n p_{pq}\right) \,.
\end{equation}
Therefore, the solution for $\partial_i p^L_j$ satisfies two conditions at any order in perturbations: (i) it is linear in $p_{ij}^{TT}$ and (ii) it is of zero order in powers of momenta in the Fourier space.

The solution of the $\pi = g_{ij} \pi^{ij}$ constraint can be cast in the following way
\begin{equation}
 \epsilon \left( p^T + \partial_i p^L_i \right) =
  - \epsilon^2 \left. h_{ij} p_{ij}\right|_{\mbox{\tiny lower order sol.}} \,.
\end{equation}
In the left-hand side one must substitute the solution for $\partial_i p^L_i$ of the same order of $p^T$ that is obtained from (\ref{eqpl}). Thus, the solution for $p^T$ from the $\pi = 0$ constraint satisfies the same two conditions of $\partial_i p^L_j$ at any order in perturbations.

Therefore, when the solutions for $\partial_i p^L_j$ and $p^T$ are inserted into the $\mathcal{H}$ and $\mathcal{C}$ constraints these remain of quadratic order in $p^{TT}_{ij}$ and the power in the momentum of the Fourier space is neither increased nor lowered.

\section{Trivial reformulation as a system with only first-class constraints}
\label{app:firstclass}

The fact that the matrix of brackets between second-class constraints acquires a triangular form suggests that this theory could be reformulated as a theory with only first-class constraints, that is, a theory with enhanced gauge symmetries. There are cases in which this procedure leads to an interesting reformulation of the original theory \cite{Gianvittorio:1991bh}. In this appendix we study this possibility, showing eventually that this procedure for the linearized theory leads to a trivial reformulation of the already known reduced theory.

The scenario is the following: since $P_N$ and $\pi$ have vanishing brackets between themselves, they could play the role of first-class constraints whereas $\mathcal{H}$ and $\mathcal{C}$ could be regarded as gauge-fixing conditions for the associated gauge symmetries. This approach requires that the constraints that are going to be promoted to first class acquire (weakly) vanishing Poisson brackets with some convenient Hamiltonian. In the case of $P_N$ and $\pi$ their brackets with the original Hamiltonian yield the other second-class constraints $\mathcal{H}$ and $\mathcal{C}$. To achieve the required condition one may add to the original Hamiltonian terms proportional to $\mathcal{H}$ and $\mathcal{C}$ such that they cancel the brackets between $P_N$ and $\pi$ and the Hamiltonian. Since the added terms vanish under $\mathcal{H} = \mathcal{C} = 0$, the interpretation is that the modified theory, which is a gauge theory, under the gauge $\mathcal{H} = \mathcal{C} = 0$ coincides with the original theory, thus both theories are physically equivalent. 

In the linearized theory the $\pi = 0$ constraint becomes also a constraint solely in the momenta, $p^T = 0$ (with $p^L = 0$ due to the momentum constraint). Notation simplifies if we also group the two linearized constraints $p_n$ and $p^T$ into a vector of momenta,
\begin{equation}
p = \left( \begin{array}{c}
p^T \\
p_n
\end{array} \right) \,.
\end{equation}
In this notation the linearized $\mathcal{H}$ and $\mathcal{C}$ constraints are $\mathbb{M} \phi$. The part of the original linearized Hamiltonian that is relevant for the present discussion is
\begin{equation}
H = \int d^3x \left(
\mathcal{H}_{\mbox{\tiny RED}} + \phi^t \mathbb{M} \phi \right) \,.
\end{equation}
The bracket between the momenta $p$ and $H$ is $ \{ p , H \} = - 2 \mathbb{M} \phi$. To get a vanishing bracket the Hamiltonian must be modified with a term proportional to the $\mathbb{M}\phi$ constraint,
\begin{equation}
\tilde{H} = H - \int d^3x\, \phi^t \mathbb{M} \phi \,, 
\end{equation}
but this subtraction leads precisely to the reduced Hamiltonian, which does not depend on $\phi$. Therefore, this procedure leads to a trivial reformulation of the linear reduced theory: the reduced theory trivially possesses the gauge symmetry generated by $p^T$ and $p_n$ since they generate full redefinitions of $h^T$ and $n$ and the reduced theory does not depend on these variables.

Trying to apply this procedure to the exact theory is much more difficult due to the involved dependence the constraints $\mathcal{H}$ and $\mathcal{C}$ have in the fields.


\end{document}